\documentclass[%
 reprint,
superscriptaddress,
 amsmath,amssymb,
 aps,
floatfix,
nofootinbib]{revtex4-1}
\usepackage{slashed}
\usepackage{subfigure,graphicx}
\usepackage{bm}
\usepackage{amsfonts}
\usepackage{xcolor}
\usepackage{amscd}

\def \Lag{\mathcal L}
\def \Act{\mathcal A}
\newcommand{\bwt}{\begin{widetext}}
\newcommand{\ewt}{\end{widetext}}
\newcommand{\newc}{\newcommand}
\newc{\beq}{\begin{equation}}
\newc{\eeq}{\end{equation}}
\newc{\beqa}{\begin{eqnarray}}
\newc{\eeqa}{\end{eqnarray}}
\newc{\nonr}{\nonumber}
\newc{\bi}{\begin{itemize}}
\newc{\ei}{\end{itemize}}
\newc{\ra}{\rightarrow}
\newc{\la}{\leftarrow}
\newc{\lra}{\longrightarrow}
\newc{\lla}{\longleftarrow}
\newc{\Lra}{\Longrightarrow}
\newc{\Lla}{\Longleftarrow}
\newc{\half}{\frac{1}{2}}
\newc{\del}{\delta}
\newc{\Del}{\Delta}
\newc{\eps}{\epsilon}
\newc{\gm}{\gamma}
\newc{\lam}{\lambda}
\newc{\kap}{\kappa}
\newc{\tri}{\triangle}
\newc{\hc}{\dagger}
\newc{\pd}{\partial}
\newc{\wt}{\widetilde}
\newc{\ovl}{\overline}
\newc{\p}{\partial}
\newc{\tchi}{\tilde{\chi}}
\newc{\ds}{\displaystyle}
\newc{\PL}{\hat{L}}
\newc{\PR}{\hat{R}}
\newc{\st}{s_\theta}
\newc{\ct}{c_\theta}

\newcommand{\eul}{\mathfrak{l}}
\newcommand{\Uel}{U(1)_\eul}
\newc{\msm}{\mathrm{SM}}
\newc{\mtev}{\mathrm{TeV}}
\newc{\clbl}{\color{blue}}
\newc{\clg}{\color{green}}
\newc{\clr}{\color{red}}

\mathchardef\mhyphen="2D

\begin{document}
\title{Study of Gauged Lepton Symmetry Signatures at Colliders}
\author{We-Fu Chang}
\affiliation{%
Department of Physics, National Tsing Hua University, Hsinchu 30013, Taiwan
}%
\affiliation{Theory Group, TRIUMF, Vancouver, B.C. V6T2A3, Canada}
\author{John N. Ng }
\affiliation{Theory Group, TRIUMF, Vancouver, B.C. V6T2A3, Canada}
\date{\today}

\begin{abstract}
We construct a new gauged $U(1)_\ell$  lepton number model which is anomaly-free for each SM generation.
The active neutrino masses are radiatively generated with a minimal scalar sector.
The phenomenology and collider signals are studied.  The interference effects among the new gauge boson, $Z_\ell$, photon, and $Z$-boson can be probed at the future $e^+e^-$ colliders even if the center-of-mass energy is below the mass of $Z_\ell$. Moreover, the electroweak precision sets a stringent bound on the mass splitting of the new lepton doublets.
\end{abstract}
\maketitle

\section{Introduction}

It is well known that the standard model (SM) Lagrangian has an accidental global symmetry associated
with the conservation of total lepton number. Equally well known is that the minimal SM cannot
accommodate the evidence of active neutrino masses from the neutrino oscillations data. If one
allows the  dimension-5 Weinberg operator (Wo)\cite{WO} of the form $\frac{c}{\Lambda} L L H H$ where
$L$ is the lefthanded lepton doublet, $ H$ denotes the Higgs field, $\Lambda$ is an unknown
high scale and $c$ is a free parameter, then after spontaneous symmetry where $H$ takes on
a vacuum expectation value $v\simeq 247$ GeV, we get a neutrino mass $m_\nu \sim \frac{c v^2}{\Lambda}$.
Since data indicate that $m_\nu \lesssim 1$ eV, the scale $\Lambda$ can range from 1 to $ 10^{11}$ TeV
depending on the value of $c$. If the neutrinos masses do indeed originate from the Weinberg
operator, it fortifies the view that the SM is an effective field theory with a small violation
of total lepton number in the form of the nonrenormalizable Wo.

The Wo gives an elegant explanation for neutrinos masses within the SM. However, its origin is not
known and is the subject of the vast field of neutrino mass models. Furthermore, whether the lepton number
is a global symmetry or a gauged symmetry and how the symmetry is broken are both open questions.
The answers or even partial answers to these questions will add immensely to our understanding of
fundamental physics. The simplest way to extend the SM and obtain the Wo is to have two or more
SM singlet righthanded (RH) neutrinos $N_R$. These singlet neutrinos can be given
heavy Majorana mass terms that change two units of lepton number explicitly by hand. Integrating
these fields out will yield the Wo at low energies, which is the well known type-I seesaw mechanism\cite{ Glashow:1979nm, *Yanagida:1979as,*Mohapatra:1979ia,*Schechter:1981cv}.
Instead of adding Majorana masses by hand, it is theoretically and phenomenologically more interesting
to generate them by the spontaneous symmetry breaking (SSB) mechanism. To this end, one adds
a SM singlet scalar field $\Phi$ and form the term $\Phi \overline{N^c_R} N_R$. When $\Phi$ gets a vacuum expectation value (VEV) $\langle \Phi \rangle \gg v$, one again gets the Type-I seesaw. If the lepton symmetry that is broken is
a $U(1)$ global symmetry, then a singlet scalar Majoron will exist in the physical spectrum and can
act as extra dark radiation \cite{CNJW}. An extended model with a dark matter candidate  has also
been constructed in \cite {CN1} and \cite {CN2}. Moreover, this symmetry can also be a local gauge symmetry.

The study of the lepton number being a local gauge symmetry has long history. If the symmetry is unbroken
one would have a leptonic photon \cite{LY}. The corresponding long range force can be searched for
in equivalence principle tests \cite{Okun}, and the limit on the leptonic fine structure constant
is $\alpha_l < 10^{-49}$. However, a complete and consistent model was not studied until
recently in \cite{FW} in conjunction with gauged baryon number.  Gauging lepton number only is given in
\cite{ST}, where active neutrino masses are given by the usual type-I seesaw model.  A different implementation with type-II seesaw mechanism\cite{Magg:1980ut, *Lazarides:1980nt, *Mohapatra:1980yp, *Cheng:1980qt} is given in \cite{Chao}. Also there the
emphasis is on constructing a consistent dark matter model with a gauged lepton number.
 More recently, a gauged $SU(2)_\ell$ model was considered by\cite{SU2_lep} with an emphasis on producing a
 dark matter candidate and baryogenesis.

In this work, we study a model of gauged lepton number $\Uel$ without employing type-I seesaw
mechanism for active neutrino masses. Specifically, the model does not have SM singlet neutrinos.
The SM leptons are assigned lepton number $\ell=1$, and $\Uel$ is spontaneously broken.
The existence of lepton specific gauge boson $Z_\eul$ is a robust prediction of
this class of models. The SM is anomalous under $\Uel$, and hence
new chiral fermions  will have to be added. Our solution differs from that of Ref.(\cite{ST})
where they use a set a fermions to solve the anomalies from all three SM lepton families
together. We choose to solve the anomaly of the SM leptons within each family, and we do not have
SM singlet neutrinos as mentioned before.

It is well known that, given two $U(1)$ gauge symmetries their corresponding gauge bosons
can have kinetic mixing \cite{Holdoma,*Holdomb} as well as mass mixing. Both are expected to be small.
The phenomenology of a kinetically mixed $Z^\prime$ with $U(1)_Y$ gauge boson was given in
\cite{CNJW2}. In this paper, we shall neglect these mixings.

This paper is organized as follows. In Sec.II, we discuss  the anomalies cancelation solution and
the new chiral fermions. Sec.III constructs the Yukawa interactions and the minimal set of scalars
required. The scalar potential that leads to symmetry breaking and the charged lepton masses of the model are also constructed and studied.
The extended gauge interactions are described in Sec IV. Particular attention is given to $Z_\eul$  which must exist in these models
 independent of which solution to the anomalies one adopts.
It is natural to assume that all the SM charged leptons carry one unit of lepton number.
Moreover, the rich phenomenology of $Z_l$ at the past and future lepton colliders is guaranteed.
 Even if its mass  denoted by $M_X$  is too heavy to be produced at these colliders, its interference with the SM $\gm$ and $Z$ can be detected in precision measurements.
 Such effects are proportional to $\frac{1}{M_{X}^{2}}$ and thus
 sensitive to low mass $Z_\eul$. These are discussed in Sec.V.
  The production at the LHC is also given
 there.  Since $Z_\eul$ couples only to leptons and not quarks,  the search strategy will have to be different from the usual extra $Z$ boson searches.
Active neutrino masses are generated by one-loop effect and is given in Sec.VI. Since it is
not the purpose of this paper to do detail neutrino oscillation study we will only present orders of
magnitude estimates.
This is followed by a discussion of  the phenomenology of the new fermions in Sec.VII.
Our conclusions are given in Sec.VIII.

\section{ Anomalies Cancelations}

 We extend the SM gauge group by $\Uel$; explicitly, it is $G=G_{\msm}\times \Uel=SU(2)\times U(1)_Y \times \Uel$.
  The color $SU(3)$ group plays no role here and can be neglected. We define the SM leptons to have number $\ell =1$ under $\Uel$. The new anomaly coefficients for a single SM
 lepton family are
 \begin{subequations}
 \label{eq:ac}
 \beqa
 \Act_1([SU(2)]^2\Uel)&=&-1/2\,, \\
 \Act_2([U(1)_Y]^2\Uel)&=&1/2\,, \\
 \Act_3([U(1)_Y[\Uel]^2)&=&0\,, \\
 \Act_4([\Uel]^3)&=&-1\,,\\
 \Act_5(\Uel)&=&-1\,,
 \eeqa
 \end{subequations}
 where $\Act_5$ is for lepton-graviton anomaly.
  We also need to check if the SM anomalies of $\Act_6([SU(2)]^2 U(1)_Y)$,$\Act_7([U(1)_Y]^3)$ and $\Act_8(U(1)_Y)$ are canceled when new chiral leptons are introduced to cancel Eq.(\ref{eq:ac}).

 We introduce two sets of chiral leptons very similar to the SM leptons. The first set consist of an $SU(2)$ doublet and a singlet and has the eigenvalue $\ell_1$. Explicitly we write
 \beq
 \label{eq:l1}
 \begin{split}
 L_{1L}=( N_{1L} ,E_{1L}) ;\;\;&[\mathbf{2},-\frac{1}{2},\ell_1]\,, \\
 E_{1R}\phantom{=(N{1L},E_{1L})} ;\;\; &[\mathbf{1},-1,\ell_1]\,,
 \end{split}
 \eeq
 where the subscript $L(R)$ stands for left(right)-handed projections.  The square parenthesis $[...]$ denotes
 $SU(2),U(1)_Y,U_\eul$ assignments.
 A second set with right-handed
 projections but lepton number$=\ell_2$ is given by
 \beq
 \label{eq:l2}
 \begin{split}
 L_{2R}=(N_{2R} ,E_{2R}) ;\;\; &[\mathbf{2},-\frac{1}{2},\ell_2]\,, \\
 E_{2L}\phantom{= (N_{2R} ,E_{2R})} ;\;\; &[\mathbf{1},-1,\ell_2]\,.
  \end{split}
 \eeq
It is easy to see that Eqs.(\ref{eq:ac}) become
\begin{subequations}
\beqa
\Act_1&=& -\frac{1}{2}(\ell_1-\ell_2 +1)\,,\\
\Act_2&=&\frac{1}{2}(\ell_1-\ell_2+1)\,,\\
\Act_3&=&0\,,\\
\Act_4&=& -\ell_1^{3}+\ell_2^{3} -1\,,\\
\Act_5&=& -(\ell_1-\ell_2 +1)\,.
\eeqa
\end{subequations}
$\Act_{1,2,5}=0$  for $\ell_2=\ell_1 +1$. Substituting into $\Act_4 =0$ gives
\beq
\ell_1(\ell_1+1)=0\,.
\eeq
The two solutions are
\begin{itemize}
\item Solution I
\beq
\label{eq:sola}
\ell_1= -1 \;\;\mathrm{and}\;\;\; \ell_2=0\,,
\eeq
\item Solution II
\beq
\label{eq:solb}
\ell_1=0 \;\; \mathrm{and}\;\;\; \ell_2=1\,.
\eeq
\end{itemize}
It is easy to check that the solutions do not contribute to $\Act_{6,7,8}$. This is not surprising
since both Eqs.(\ref{eq:l1},\ref{eq:l2})  form  vectorlike pairs under $G_{\msm}$.
Thus, Eqs.(\ref{eq:sola},\ref{eq:solb})  are anomaly-free without the use of singlet RH neutrinos.
\section{Yukawa Interactions}
After determining the anomaly-free lepton representations, we can proceed to construct $G$-invariant Yukawa interactions. This will produce the minimal scalar fields required for viable charged and neutral lepton mass matrices at the tree level.

We will give a detail discussion of the physics associated with solution (I) \footnote{Solution II gives qualitatively the same physics. It is easy to extend our discussions to this case.}.
The complete set of leptons for this solution and their gauge quantum numbers are given in Table(\ref{tb:lA}).
With this, one can form all possible Lorentz-invariant bilepton combinations that are invariant under
$G_\msm$. The next step is to identify scalar fields that will make Yukawa interactions that
are invariant under the full gauge group $G$. Besides the SM Higgs field $H_0$, the minimum set of new scalars we require are  $\Phi,S$, and $H_1$ and
their quantum numbers are also given in Table(\ref{tb:sA}).
\begin{table}
\begin{center}
\renewcommand{\arraystretch}{1.30}
\begin{tabular}{|c|c|c|c|}
\hline
Field&$SU(2)$&$\phantom{U}Y\;\;$&$\phantom{U}\ell\;\;$\\ \hline
$\ell_L=\begin{pmatrix} \nu_L\\e_L \end{pmatrix}$ &{\bf{2}}&$-\frac{1}{2}$&$\phantom{-}1$\\ \hline
$e_R$&{\bf{1}}&$-1$&$\phantom{-}1$\\ \hline
$L_{1L}=\begin{pmatrix}N_{1L}\\E_{1L}\end{pmatrix}$&\bf{2}&$-\frac{1}{2}$&$-1$\\ \hline
$E_{1R}$&{\bf{1}}&$-1$&$-1$ \\ \hline
$L_{2R}=\begin{pmatrix}N_{2R}\\E_{2R}\end{pmatrix}$&{\bf{2}}&$-\frac{1}{2}$&$\phantom{-}0 $\\ \hline
$E_{2L}$&{\bf{1}}&$-1$&$\phantom{-}0 $\\ \hline
\end{tabular}
\caption{Lepton fields for anomaly-free solution I}
\label{tb:lA}
\end{center}
\end{table}

With this, the Yukawa interactions are given by\footnote{ We are interested in the minimal setup. In general there could also be a Yukawa term $ \ovl{L}_{1L}e_R H_2$ with a second doublet $H_2:(2,1/2,-2)$.}
\beq
\label{eq:YA}
\begin{split}
\Lag_y=&
y_e\ovl{\ell_L}e_R H_0   + Y_2 \ovl{L_{1L}}E_{1R} H_0 +
 Y_3\ovl{L_{2R}}E_{2L} H_0 \\
 &+\lam_1 \ovl{\ell_L} L_{2R}\Phi_1 +\lam_2 \ovl{E_{2L}}e_R \Phi^\dagger_1 +
 \lam_3\ovl{L_{1L}}L_{2R}\Phi_1^* \\
 &+\lam_4 \ovl{E_{2L}}E_{1R}\Phi_1
 + Y_1 \ovl{\ell_L}H_1 E_{1R} + f\ovl{\ell_L^c}\eps L_{1L}S  \\
 &+h.c\,.
\end{split}
\eeq
\begin{table}
\begin{center}
\renewcommand{\arraystretch}{1.30}
\begin{tabular}{|c|c|c|c|}
\hline
Field&$SU(2)$&$\phantom{U}Y\;\;$&$\phantom{U}\ell \;\;$\\ \hline
$H_0=\begin{pmatrix}0\\ \frac{v+h}{\sqrt{2}}\end{pmatrix}$ &{\bf{2}}& $\frac{1}{2}$&0 \\ \hline
$H_1=\begin{pmatrix} H_{1}^{+}\\ H^{0}_1\end{pmatrix} $ & {\bf{2}} & $\frac{1}{2}$& 2\\ \hline
$S$ & {\bf{1}}&1&0 \\ \hline
$\Phi_1$ & {\bf{1}}& 0 &1 \\ \hline
$\Phi_2$ & {\bf{1}}&0& 2 \\ \hline
\end{tabular}
\caption{ Minimal scalar fields for leptons of solution I}
\label{tb:sA}
\end{center}
\end{table}
\vspace{0.5cm}

It can be seen that $S$  is charged and cannot develop a VEV. $H_{1}$ is a Higgs-like field and may or may not pick up a VEV depending  on the parameters in the scalar potential. Here, we make the reasonable
assumption that the lepton-number breaking scale is much higher than $v$. In order not to have a weak-scale lepton-number violation, we will work in the parameter space where $H_1$ is not Higgssed
since it has $\ell=2$. Moreover, $\Phi_1$ is a
neutral scalar, and it can pick up a VEV $w$ and thus can bestow masses to the new charged leptons $E_{1,2}$. They will be much heavier than SM charged leptons if $w\gg v$. Another electroweak singlet scalar $\Phi_2$ with  2 units of lepton number is required for neutrino mass generation as we shall see later. But it does not enter in Eq.(\ref{eq:YA}). We will not include scalar fields with $|Y|>1$ as they play no role in our study.

Having specified all the necessary scalars, the minimal $G$-invariant scalar potential is given by
\beq
\label{eq:VA}
\begin{split}
&V(H_0,H_1,\Phi_1,\Phi_2,S)=\\
&-\mu^2 H_{0}^\dagger H_0  + m_1^{2}H_1^{\dagger}H_1+\kap_0\left(H_0^\dagger H_0\right)^2+\kap_1\left( H_1^\dagger H_1\right)^2 \\
&+ \kap_2\left(H_0^\dagger H_0\right)\left(H_1^\dagger H_1\right)+\kap_3\left(H_0^\dagger H_1\right)\left( H_1^\dagger H_0\right)\\
&+m_S^2 S^\dagger S +\kap_S\left(S^\dagger S\right)^2-\sum_{i=1,2}\mu_{i}^2 \Phi_i^\dagger\Phi_i \\
&+\kap_{11}\left(\Phi_1^\dagger\Phi_1\right)\left(\Phi_1^\dagger \Phi_1 \right)+\kap_{12}\left(\Phi_1^\dagger\Phi_1\right)\left(\Phi_2^\dagger \Phi_2\right) \\
&+\kap_{22}\left(\Phi_2^\dagger\Phi_2\right)\left(\Phi_2^\dagger \Phi_2 \right)\\
&+\sum_{i=1,2}\sum_{j=0,1}\kap_{\Phi_i H_j}(\Phi_i^\dagger\Phi_i)(H^{\dagger}_j H_j)\\
&+\sum_{i=1,2}\kap_{\Phi_i S}(\Phi_i^\dagger\Phi_i)(S^\dagger S)+\sum_{i=0,1,2}\kap_{H_i S}(H_i^{\dagger} H_i)(S^\dagger S)\\
&+\lam_{1\ell} H_1 \eps H_0 S^\dagger\Phi_2^\dagger + \lam_{2\ell}H_0^\dagger H_1  (\Phi_1^*)^2 \\
&+ \mu_3 H_0^\dagger H_1 \Phi_2^* +\mu_4 (\Phi_1^*)^2\Phi_2 +h.c.
\end{split}
\eeq
Lepton-number violation occurs spontaneously for $\langle|\Phi_{1,2}|\rangle =w/\sqrt{2} \neq 0$
and is the only such scale in the model\footnote{In general $\Phi_1$ and $\Phi_2$ need not have the same VEV. This only adds more parameters to the model without adding more physics. We shall assume they are equal.}. Thus, we write $\Phi_{1(2)}=\frac{w+\varphi_{1,(2)}}{\sqrt{2}}$.

After SSB, the lepton mass matrices arise  from Eq.(\ref{eq:YA}).
The  charged lepton matrix in the basis $ \mathcal{E}=(e_w,E_{1},E_2)$\footnote{Here, we introduce the intermediary subscript $w$ to $e$ to remind us it is the weak basis.}
is
\beq
\label{eq:CLMA}
M_E=\frac{w}{\sqrt{2}}\begin{pmatrix}y_e r &0 & \lam_2\\0& Y_2 r & \lam_4\\ \lam_1 & \lam_3 &Y_3 r \end{pmatrix}
\eeq
where $r \equiv \frac{v}{w} \ll 1$.
And the neutral lepton mass matrix in terms of the chiral
states $(\nu_L^{w}, N_{1L}, N_{2R}^c)$\footnote{Again, the intermediary superscript  $w$ is introduced for neutrino.}
is
\beq
\label{eq:NMA}
M_N=\frac{w}{\sqrt{2}}\begin{pmatrix}0&0& \lam_1 \\0&0&\lam_3 \\ \lam_1 & \lam_3 &0
\end{pmatrix}\,.
\eeq
The identity $\bar{\psi}^c_1(\hat{L}/\hat{R}) \psi_2^c= \bar{\psi}_2(\hat{L}/\hat{R}) \psi_1$ has been used to give the symmetric $M_N$, which is a tree-level
result. At this level, the active neutrino is massless.

However, what is interesting is that Eq.(\ref{eq:VA}) has sufficient structure to give a one-loop
radiative Majorana mass to $\nu_L^{w}$; i.e. the upper leftmost entry of Eq.(\ref{eq:NMA}) will have a quantum contribution.  The source comes from  the term involving  $\lam_{2\ell}$, which spontaneously breaks lepton symmetry when $\Phi_2$ gets a VEV. It also induces a mixing between the charged scalars $H_{1}^{ +}$ and $S^+$. The details of radiatively generated active neutrino masses will be discussed in a later section.

 Eq.(\ref{eq:YA}) holds for a single lepton family. It can be generalized to the three-families case by promoting
 the couplings $y_e, Y_{1,2,3},\lam_{1,\dots,4},f$ to $3\times 3$ matrices. There are also similar terms
 connecting different families, which we  will neglect since we are not interested in charged lepton flavor violation or flavor changing neutral current processes in this paper. Henceforth, our discussions
 will mostly involve only a single lepton family which is designated as the electron family.

\subsection{A quartet of scalar fields}
 It is easy to see from Eq.(\ref{eq:VA}) that the SM Higgs field in the gauge basis can be identified with $H_0$.
The SM Higgs field will mix with the real parts of the SM singlets $ \Re \Phi_{1,2}$ and  SM doublet  charge neutral part $\Re H_1^0$ through the quartic couplings
$\lam_{2\ell}$, $ \lam_{\Phi_1 H_0}, \lam_{\Phi_2 H_0}$ and the cubic term $\mu_3$ after SSB. The scalar mass matrix is in general $4\times 4$.
We denote this quartet of gauge states by $\mathcal{H}_\alpha =\frac{1}{\sqrt{2}}( \Re H_0^0,\;\Re H_1^0,\; \Re \Phi_1,\; \Re \Phi_2)$.
As usual, the mass eigenstates $\mathfrak{h}_i=(h_0,\; h_1,\; h_2,\;h_3 )$ are related to $\mathcal{H}$ via $\mathfrak{h}= U \mathcal{H}$ where $U$ is a $4\times 4$
 unitary mixing matrix. The strength of the mixings given by the elements $U_{i\alpha}$ will depend on the physical masses of the new scalars and the quartic couplings. Since no beyond-the-SM
scalars are found at the LHC, we make the conservative assumption that they are all heavier than 800 GeV. However, we are mindful that optimal search strategy for a specific scalar is model dependent. Nevertheless, a robust prediction is that a  universal suppression factor $U_{11}$ applies to all SM Higgs couplings which can be probed by the Higgs signal strengths at the LHC. The SM signal strength is parameterized by $\mu$ and is unity for the SM.  The LHC-1 bound is
$\mu=1.09\pm 0.11$ \cite{LHCmu}. This implies $|U_{11}|^2 > .87$ at $2\sigma$ level. Hence, the mixing of $H_0$ with any of the other three scalars must be quite small or even vanishing. Small mixings  can be achieved by tuning the
couplings $\lam_{2\ell}$,  $ \lam_{\Phi_1 H_0}, \lam_{\Phi_2 H_0}$, and $\mu_3$.

\subsection{Two simplified cases of lepton mass diagonalization }
To capture the physics essence of this model, we will avoid the complication of keeping track of all the free parameters and
focus on two simplified scenarios:
\begin{itemize}
\item
 \textbf{Scenario-A}: We take  $\lam_{2\ell}$ and $\mu_3$ to be small but finite, and
  we also assume that  $Y_2 \sim Y_3 \sim Y$ and  $\lam_{1,2,3,4}\sim \bar{\lam}$.
  \item
 \textbf{Scenario-B}: This is the limiting case of
   $\lambda_{1,2}=0=\mu_3$ . We also assume that $Y_2 \sim Y_3 \sim Y$ and  $\lam_{3,4}\sim \bar{\lam}$.
\end{itemize}
We shall refer to them as the Yukawa symmetry limits.

\subsubsection{ Scenario A}

From Eq.(\ref{eq:YA}), the charged lepton matrix in the basis $ \mathcal{E}=(e_w,E_{1},E_2)$ is
\beq
\label{eq:CLMA_A}
M_E\sim \frac{w}{\sqrt{2}}\begin{pmatrix}y_e r &0 &\bar{\lam}\\0& Y r &\bar{\lam}\\ \bar{\lam}&\bar{\lam}&Yr \end{pmatrix}\,.
\eeq
In the limit $r\ra 0$ and $y,\bar{\lam}$ finite, the charged lepton mass matrix has the structure
\beq
\begin{pmatrix} 0&0&1\\0&0&1\\1&1&0
\end{pmatrix}\,.
\eeq
The spectrum consists of a massless electron and two heavy degenerate leptons with mass  $\sim \bar{\lam} w$  in this limit.
Returning to Eq.(\ref{eq:CLMA_A}), it can be shown that the smallest eigenvalue is given by the larger of $y_e,Y$. This implies that in order to get  the electron mass right, $Y\simeq y_e$. Thus, without loss
of generality we write the charged lepton mass matrix as
\beq
\label{eq:clm}
M_E \simeq \begin{pmatrix} m_e&0&\bar{\lam} w/\sqrt{2} \\0& m_e &\bar{\lam} w/\sqrt{2} \\ \bar{\lam} w/\sqrt{2} & \bar{\lam} w/\sqrt{2} & m_e
\end{pmatrix}\,,
\eeq
where $m_e$ is the physical electron mass. Moreover, the parameter $Y_{1}$ remains free.

In general, we can write the physical mass eigenstates $\mathcal{E}_\alpha^\prime=(e,E_-,E_+)$ where $\alpha=1,2,3$ is given by
\beq
\label{eq:DMe}
\mathcal{E}_i^\prime=V_{i \alpha } \mathcal{E}_\alpha\,,
\eeq
 where $V$ is the unitary matrix that diagonalizes
$M_E$ such that  $(V^A)^T\cdot M_E \cdot V^A = \mbox{diag}\{ m_e, -(\bar{\lam} w -m_e), \bar{\lam} w+ m_e\}$.  For the simplified symmetrical case of
Eq.(\ref{eq:clm}), $V^A$ can be worked out to be
\beq
\label{eq:Usym_A}
V^A\simeq \left(
    \begin{array}{ccc} \frac{1}{\sqrt{2}}&\frac{1}{2} & \frac{1}{2} \\
     -\frac{1}{\sqrt{2}}&\frac{1}{2} & \frac{1}{2} \\
     0& -\frac{1}{\sqrt{2}}&\frac{1}{\sqrt{2}} \\
    \end{array}
  \right)\,.
\eeq

The neutral lepton mass matrix is
\beq
\label{eq:NMA_A}
M_N\simeq
\frac{\bar{\lam}w}{\sqrt{2}}\begin{pmatrix}0&0& 1\\0&0&1\\ 1&1&0
\end{pmatrix}\,.
\eeq
 Note that the difference between Eqs.(\ref{eq:NMA}) and (\ref{eq:clm}) is proportional to an identical matrix.
 Therefore, both the neutral and charged lepton mass matrices are diagonalized by the same rotation, Eq.(\ref{eq:Usym_A}).
At tree level, the spectrum consists of a massless neutrino and a Dirac neutrino of mass $\sim \bar{\lam} w$. This can be seen by defining $n\mp= \frac{1}{\sqrt{2}}(\nu^w_L\mp N_{1L})$. In the basis $( n_{-},n_{+},N_{2R}^c )$,
the matrix $M_N$ becomes
\beq
M_N \propto
\begin{pmatrix} 0&0&0\\0&0&1\\0&1&0 \end{pmatrix}.
\label{eq:numm}
\eeq
 Clearly, $n_{-}$ is massless and the pair of Weyl neutrinos $n_{+},N_{2R}^c$ combines into a Dirac neutrino.
In the case that the neutrinos receive notable  quantum corrections, we denote
the charged neutral mass eigenstates as $(\nu, N_-,N_+)$ with the convention $M_{N_+}> M_{N_-}$.

The $Y_1$ Yukawa term is relevant when considering the exotic fermion decays.
In the mass basis, we have the following terms:
\beq
\begin{split}
\frac{Y_1}{2}& \left\{- \bar{e}[ \Re(H_1^0) + i \gamma_5 \Im(H_1^0) ] e \right. \\
  &\left.+ \frac{1}{\sqrt{2}} \bar{e} [\gamma_5 \Re(H_1^0) + i \Im(H_1^0) ] (E_+ + E_-)\right.\\
 &+ \left.\frac{1}{2}(\bar{E}_+ +\bar{E}_{-})[\Re(H_1^0)+i\gm_5 \Im(H_1^0)](E_+ +E_{-}) \right\} \\
&-\frac{Y_1}{2} \left[\bar{\nu}+ (\bar{N}_++\bar{N}_-)/\sqrt{2} \right] \hat{R}\, e H_1^+\\
&+\frac{Y_1}{2\sqrt{2}} \left[\bar{\nu}+ (\bar{N}_++\bar{N}_-)/\sqrt{2} \right] \hat{R}\, (E_+ +E_-) H_1^+ \\
&+h.c.
\end{split}
 \label{eq:H1Yukawa_A}
\eeq

\subsubsection{ Scenario B}
In this case, the SM leptons completely decouple from the exotic fermion sector.
The lepton matrices now become
\beq
\label{eq:CLMA_B}
M_E\sim \frac{w}{\sqrt{2}}\begin{pmatrix}y_e r &0 &0 \\0& Y r &\bar{\lam}\\ 0&\bar{\lam}&Yr \end{pmatrix}\,,
\eeq
and $M_E$  can be diagonalized by the rotation matrix
\beq
\label{eq:UsymB}
V^B\simeq \left(
    \begin{array}{ccc} 1&0&0\\
    0&\frac{1}{\sqrt{2}} & \frac{1}{\sqrt{2}}  \\
    0& -\frac{1}{\sqrt{2}}&\frac{1}{\sqrt{2}} \\
    \end{array}
  \right)\,.
\eeq
The mass eigenstates are again denoted as $(e,E_-,E_+)$

For neutrinos,
\beq
\label{eq:NMA_B}
M_N =
\frac{w \lam_3}{\sqrt{2}}\begin{pmatrix}0&0& 0 \\0&0&1 \\ 0&1&0
\end{pmatrix}\,.
\eeq
Clearly, there is no mixing between the $\nu^w$ and $(N_{1L}, N_{2R}^c)$.

In the chiral basis, $M_N$ is also diagonalized by $V^B$, and the mass eigenstates are again denoted as $(\nu, N_-,N_+)$.
At tree level, $N_+,N_-$ are degenerated.
In fact, $N_{1L}$ and $N_{2R}$ form a Dirac fermion at tree-level ( let us simply call it $N$)
and $\hat{L}N=N_{1L}, \hat{R}N=N_{2R}$.
 The degeneracy will be broken by the one-loop mass correction.
However, the quantum correction is expected to be much smaller than $w$, and taking $N$ as a Dirac DM is a good approximation.

The $Y_1$ Yukawa term
in the mass basis becomes
\beqa
 \frac{Y_1}{\sqrt{2}} ( \bar{\nu} H_1^+ +\bar{e} H_1^0) \hat{R}(E_+ + E_-) +h.c.
 \label{eq:H1Yukawa_B}
\eeqa
The heavier of $E_\pm$ and $H_1$ can decay via this Yukawa interaction followed by the lighter one decaying through gauge interactions.

\section{Gauge Interactions}

The covariant derivative is
\beq
\label{eq:cod}
D_\mu= \pd_\mu -i\frac{g}{2} \mathbf{W}_\mu\cdot \boldsymbol{\tau}-ig^\prime Y B_\mu -ig_\eul (\ell)Z_{\eul \mu}\,,
\eeq
where $Z_\eul$ is the gauge boson for $\Uel$, $Y$ the hypercharge, and $\ell$ the lepton number.
All the quantum numbers can be read from the tables. Other notations are standard.
 After the SSB of $\Uel$, $\langle\Phi_{1,2}\rangle=w_{1,2}$, the $Z_\eul$ acquires a mass
\beq
\begin{split}
M_X&= g_\eul \sqrt{w_1^2+4 w_2^2}\\
&= g_\eul \bar{w}\\
&= 2.24 g_\eul w \;\; {\text{for}} w_1=w_2=w\,,
\end{split}
\label{eq:MX}
\eeq
where $\bar{w}^2=w_1^2 +4 w_2^2 $ gives the overall lepton number violating scale.

In terms of the physical gauge bosons, the gauge interaction in the weak basis is
\beqa
&&i e \left[\bar{e}_w\gamma^\mu e_w+ \bar{E}_1\gamma^\mu E_1 +\bar{E}_2\gamma^\mu  E_2 \right]P_\mu \nonr\\
&-&\frac{i g_2}{2 c_w} \left[\bar{\nu}_w\gamma^\mu\hat{L} \nu_w+ \bar{N}_1\gamma^\mu\hat{L} N_1 +\bar{N}_2\gamma^\mu\hat{R} N_2 \right]Z_\mu\nonr\\
&-&\frac{i g_2}{c_w} \left[\bar{ e}_w\gamma^\mu(g_L \hat{L}+g_R \hat{R})e_w
+ \bar{ E}_1\gamma^\mu(g_L \hat{L}+g_R \hat{R})E_1 \right. \nonr \\
&&\left. \quad\quad +\bar{E}_2\gamma^\mu(g_L \hat{R}+g_R \hat{L}) E_2
\right]Z_\mu\nonr\\
&-&\frac{i g_2}{\sqrt{2}} \left[\bar{\nu}_w\gamma^\mu\hat{L}e_w+ \bar{N}_1\gamma^\mu\hat{L} E_1 +\bar{N}_2\gamma^\mu\hat{R}  E_2 \right]W^+_\mu + h.c.\nonr\\
&-& i g_\eul \left[\bar{\nu}_w\gamma^\mu\hat{L}\nu_w + \bar{e}_w\gamma^\mu e_w -\bar{N}_1\gamma^\mu\hat{L} N_1 -\bar{E}_1\gamma^\mu  E_1 \right](Z_\eul)_\mu\,, \nonr \\
&&
\eeqa
where $P$ denotes the photon,
$g_L=-1/2+s_w^2$ and $g_R=s_w^2$ are the SM left-handed and right-handed $Z$-electron couplings, respectively. We have also assumed the kinetic mixing between $\Uel$ and $U(1)_Y$ is negligible.

\subsection{SM gauge interaction}
For the scenario-B, the weak basis and the mass basis are related by
\beqa
 e_w &=&e, \nonumber \\
  E_1 &=&\frac{1}{\sqrt{2}}[E_+ + E_- ], \nonumber \\
 E_2&=&\frac{1}{\sqrt{2}}[E_+ - E_- ]\,,\\
 \nu_w &=&\nu\,, \nonumber \\
 N_{1L} &=&\hat{L}N=\frac{1}{\sqrt{2}}[N_+ + N_- ]\,,\nonumber \\
 N_{2R}&=&\hat{R}N=\frac{1}{\sqrt{2}}[N_+^c - N_-^c ]\,.
\eeqa
The mass splitting between $N_+$ and $N_-$ could stem from the quantum corrections and is unlikely
to be experimentally detectable. Hence, it is a very good approximation to
lump them into a Dirac fermion $N$.

The QED interaction in the mass basis remains intact,
\beq
i e \left[\bar{e}\gamma^\mu e+ \bar{E}_+\gamma^\mu E_+ +\bar{E}_-\gamma^\mu  E_- \right]P_\mu\,.
\eeq
The SM charged current(CC) interaction becomes
\beq
-\frac{i g_2}{\sqrt{2}} \left[
\bar{\nu}\gamma^\mu\hat{L} e+ \frac{1}{\sqrt{2}} \bar{N} \gamma^\mu  E_+
- \frac{1}{\sqrt{2}} \bar{N} \gamma^\mu\gamma^5  E_-
\right]W^+_\mu + h.c.
\eeq
The SM CC interaction is intact. However, note that the $\bar{N} E_+ W^+$ vertex is vectorlike and the $\bar{N}E_- W^+$  one is axial vector.

The SM neutral current(NC) interaction admits a similar structure and becomes
\beqa
&& \left\{- \frac{i g_2}{2c_w} \left[ \bar{\nu }\gamma^\mu\hat{L} \nu + \bar{N}\gamma^\mu  N \right]-\frac{i g_2}{c_w} \left[
\bar{e}\gamma^\mu (g_L\hat{L}+g_R \hat{R}) e\right] \right .\nonr\\
&& -\frac{i g_2}{c_w}\frac{g_L+g_R}{2} \left[  \bar{E}_+ \gamma^\mu E_+
+\bar{E}_-\gamma^\mu E_- \right]\nonr\\
&& \left.-\frac{i g_2}{c_w}\frac{g_R-g_L}{2} \left[  \bar{E}_+ \gamma^\mu\gamma^5 E_-
+\bar{E}_-\gamma^\mu\gamma^5 E_+ \right] \right\}\times  Z_\mu\,.
\eeqa

For the  scenario-A,
\beqa
 e_w  &=&\frac{e}{\sqrt{2}}  +\frac{1}{2} [E_+ + E_- ] \, ,\nonr \\
E_1 &=& - \frac{e}{\sqrt{2}} +\frac{1}{2} [E_+ + E_- ]\,, \nonr \\
E_2&= &\frac{1}{\sqrt{2}}[E_+ - E_- ]\,,\\
\nu^w_L &=&\frac{\nu}{\sqrt{2}} +\frac{1}{2} [N_+ + N_- ] \,,\nonr \\
N_{1L}& =& - \frac{\nu}{\sqrt{2}}  +\frac{1}{2} [N_+ + N_- ]\,, \nonr \\
N_{2R}&=&\frac{1}{\sqrt{2}}[N_+^c - N_-^c ]\,.
\eeqa
Again, we adopt the approximation in which  $N_+, N_-$ form a Dirac fermion $N$.
The neutrinos in the interaction basis become
\beq
\nu^w_L \simeq \frac{1}{\sqrt{2}} ( \nu + \hat{L} N ) \,,
N_{1L} \simeq \frac{1}{\sqrt{2}} (-\nu + \hat{L} N )\,,
N_{2R}\simeq \hat{R} N\,.
\eeq
And it is easy to check that the QED, SM-CC, SM-NC parts are the same  as those in scenario-B.

In these two cases we discussed, the SM gauge couplings are intact. This is due to that $V_{31}=0$ for both cases.
In general, the SM $Z\mhyphen e\mhyphen e$ and $Z\mhyphen\nu\mhyphen\nu$ axial-vector part couplings can deviate from the SM prediction.
However, the deviation is expected to be small which is controlled by the mixing, $\sim {\cal O}(m_l/v_1) <10^{-7}$, between SM lepton and $L_2$.

\subsection{ $Z_\eul$ interactions}
 The corrections from any extra Z boson couplings  to SM leptons  are important for low-energy
 high-sensitivity experiments, which can be done at the proposed lepton colliders such as the ILC\cite{ILC} and CLIC\cite{CLIC}.
 And it is particularly true for $Z_\eul$. To facilitate such studies, we need to know the couplings of $Z_\eul$ to the SM leptons, which are also important for direct searches.

We begin with the couplings of charged leptons. It is easy to see from Eq.(\ref{eq:cod}) and
 Table(1),  the charged leptons
$ \mathcal{E}=(e_w,E_1,E_2)$ have vector couplings to $Z_\eul$ in the gauge basis. We define a charged matrix $Q^{\mathcal{E}}$ representing this coupling by $\ovl{\mathcal{E}}Q^{\mathcal{E}}\gm_\mu \mathcal{E}Z_\eul^\mu$, where
\beq
\label{eq:QE}
Q^{\mathcal{E}}= \begin{pmatrix}1&0&0 \\ 0&-1&0\\ 0&0&0 \end{pmatrix}\,.
\eeq
In the mass basis the corresponding charge mass is given by
\beq
\label{QEP}
Q^{\prime \mathcal{E}} = V^\dagger Q^{\mathcal{E}} V\,,
\eeq
where $V$ is given by Eq.(\ref{eq:DMe}). Specific examples of $V$ are given in Eq.(\ref{eq:Usym_A}) and Eq.(\ref{eq:UsymB}). In general, $Q^{\prime \mathcal{E}}$ is not even a diagonal matrix, and this is in sharp contrast to case of the SM gauge interactions. Of particular interest is
$Q^{\prime \mathcal{E}}_{1 1}= |V_{e 1}|^2 -|V_{E 1}|^2 $,  which determines the coupling strength of $Z_\eul$ to the physical electrons.
For scenario-A, we see that not only
is this suppressed by matrix elements but an accidental cancelation also occurs. Indeed, for the simplified case it vanishes as from  Eq.(\ref{eq:Usym_A}).

Similarly, we find  $Q^{\prime \mathcal{E}}_{1 2} =V^*_{e 1}V_{e 2}- V^*_{E 1}V_{E 2}$, which gives the off-diagonal coupling of $Z_\eul$ to the physical electron and new heavy charged lepton. Explicitly, we
have for scenario-A and in Yukawa symmetry limit
\beq
\label{eq:ZleeA}
-\frac{ i g_\eul}{\sqrt{2}}\left[ \bar{e}\gamma^\mu (E_++E_-) + (\bar{E}_++\bar{E}_-)\gamma^\mu e \right]Z_{\eul\mu}\,.
\eeq
In general, the coupling $Q^{\mathcal{E}}_{11}\bar{e}\gm_\mu e\, Z^\mu_{\eul}$ will not vanish since the $\lam$'s are all different and a complete cancelation is not expected. Moreover, it is expected that $|Q^{\mathcal{E}}|<1$.

On the other hand, it is very different for scenario-B. In this case, the SM electron decouples from the exotic fermion, and one has $Q^{\prime \mathcal{E}}_{1 1}=1$. Instead, we have
\beq
\label{ZleeB}
 ig_\eul\left[\half (\bar{E}_+ +\bar{E}_{-})\gm^\mu(E_{+}+E_{-})-\bar{e}\gm^\mu e\right]Z_{\eul \mu}\,.
\eeq
In both scenarios, the couplings are vectorial.

Similar considerations for the neutral leptons give for scenario-A
\beq
\label{eq:ZLNA}
-\frac{ i g_\eul}{2}\left[ \bar{\nu} \gamma^\mu\hat{L} N + \bar{N} \gamma^\mu\hat{L} \nu
\right]Z_{\eul\mu}
\eeq
and for scenario-B
\beq
\label{eq:ZLNB}
ig_\eul\left[ \bar{N}\gm^\mu \hat{L} N -\bar{\nu}\gm^\mu\hat{L}\nu \right] Z_{\eul\mu}\,.
\eeq
In contrast to the charged leptons, these couplings are left-handed.

\section{Phenomenology of $Z_\eul$}
It is clear that $Z_\eul$ has only tree-level couplings to SM leptons and not to quarks. Hence its phenomenology is very different from most extra Z extensions of the SM.

For scenario-A in the Yukawa symmetry limit,  $Z_\eul$ does not couple to SM charged leptons phenomenology at tree level, although one-loop effect can exist.
Thus, we do not expect such probes to be sensitive to $Z_\eul$ in this limiting case. On the other hand, for scenario-B, this coupling is at full strength.
Therefore, in the following, we mainly focus on the $Z_l$ phenomenology for scenario-B. In between the two cases, our results  can be used by properly multiplying by the appropriate factor $(Q^{\prime \mathcal{E}})^2$ once elements of the mixing matrix $V$ is determined.

The model has many parameters. However, most of them are related to the exotic scalars.
For $Z_\eul$ phenomenology, they largely do not play a role. The controlling parameters are $g_\eul$ and
 $M_X$, the mass of $Z_\eul$. Whether the new leptons and scalars are heavier or lighter than
 $Z_\eul$ mainly  affects the branching ratio of $Z_\eul$ into SM states, and is of secondary
 importance here. For definiteness, we shall assume that $Z_\eul$ is the lightest of the new
 particles.

Direct production from $e^+e^-$
colliders via $e^+e^-  \ra Z_\eul$, which subsequently decays into $\ell^+ \ell^-$ pairs, gives
 unambiguous signal if kinematically allowed. Indirect virtual exchange of $Z_\eul$ effects can
 be discerned in low-energy precision experiments involving only leptons. Some notable
 reactions are studied below.

\subsection{LEP II bound}

  The four-lepton contact interactions between electrons and charged leptons $\ell$  with scale $\Lambda_{VV}$\footnote{Note that if $l=e$ there will be an extra symmetry factor 2 in the denominator of Eq.(\ref{eq:4fermi}). } are parameterized by
\beq
\frac{4\pi}{(\Lambda_{VV})^2 } \left(\bar{e}\gamma^\mu e \right)\left(\bar{l}\gamma_\mu l\right)\,.
\label{eq:4fermi}
\eeq
This can be generated by exchanging a heavy $Z_\eul$ boson with the coupling $g_\eul$. Since the leptons
are mass eigenstates, the coupling has to be scaled by the factor $Q^{\prime\mathcal{E}}_{1 1}$.
The operator yields a destructive interference with the SM process for $\sqrt s \lll M_X$.   The effects
of the contact interactions have been searched for at LEP. A limit $\Lambda_{VV}>20.0$ TeV is set  if the universality between leptons is assumed\cite{LEP}.
This amounts to a $\rho$-dependent lower bound on $M_X$,
\beq
M_X \geq \sqrt{\rho}\sqrt{\alpha}\times 20.0\, \mbox{TeV} \sim 1.77 \sqrt{\rho}\, \mbox{TeV}\,.
\label{eq:MXbound}
\eeq
where $\rho\equiv (g_\eul/e)^2$.
For example, $M_X>0.97$ TeV if $\rho=0.3$.
The above limit works for scenario (B), in which $Q^{\prime \mathcal{E}}_{1 1}=1$.
On the other hand, there is no such tree-level contact interaction for scenario-A since $Q^{\prime \mathcal{E}}_{1 1}=0$ and the LEP bound does not apply at all.

For the remainder of this section, scenario (B) is assumed.

\subsection{$Z_\eul$ width}
With the assumptions listed, the main decay modes of  $Z_\eul$  are  into the SM leptons.
The total width can be calculated to be:
\beq
\label{eq:Zlwidth}
\Gamma_{Z_\eul} = \sum_l \frac{\alpha \rho }{6} M_X(1+2 x_l)\sqrt{1-4x_l} \left[ (l_L^l)^2+(l_R^l)^2\right]\,,
\eeq
where $x_l=(m_l/M_X)^2$,
$\alpha$ the fine structure constant, and the  $ l^l_{L,R}$  is the left-/right-handed coupling for the $l$ lepton flavor. Since $x_l \ll 1$, we have $\Gamma_{Z_\eul}= \frac{3}{2}\alpha\rho M_X $.
For a light $Z_\eul$, $M\sim {\cal O}(100)$ GeV, its typical width is around a few GeV, and its decay branching ratios are $Br(Z_\eul\ra l^+ l^-)= 2/9$ and $Br(Z_\eul\ra \nu \bar{\nu})= 1/9$  for each flavor.
\subsection {Front-back asymmetry ($A_{FB}$) in $e^+ e^- \ra \mu^+ \mu^-$}

\begin{figure*}[htb!]
 \centering
 \includegraphics[width=0.6\textwidth]{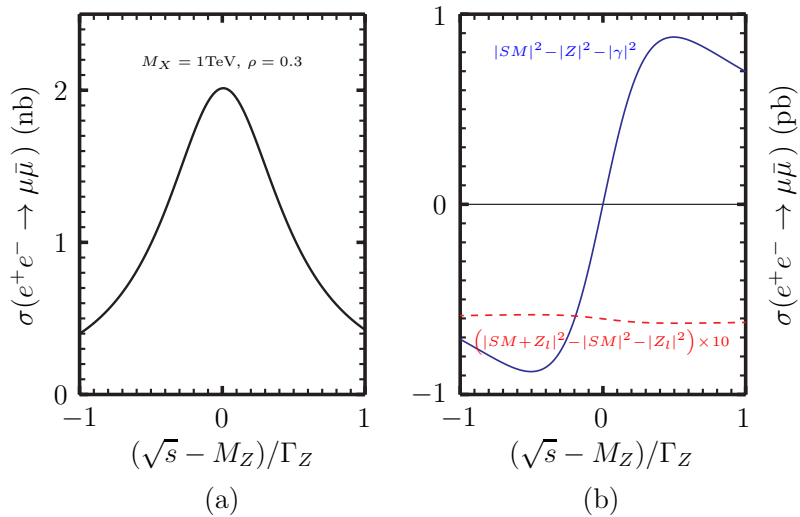}
\caption{(a) The line shape of $e^+e^-\ra \mu \bar{\mu}$ cross section near the Z-pole for $M_X=1$TeV and $\rho=0.3$.
(b) The SM photon-$Z$ interference, and the $Z_l$-SM interference. }
\label{fig:Z_LS}
 \end{figure*}

The exchange of $Z_\eul$, which has vector couplings to $e,\mu$, will interfere with the SM exchange of
$Z,\gm$. The differential cross section is given by

\beqa
\label{eq:eeff}
\frac{d \sigma}{d \cos\theta} &=&  \frac{\pi \alpha^2}{2 s}\left\{  |D_{\gamma \eul}|^2(1+\cos^2\theta)\right.\nonr\\
&&+\frac{1}{4(s_W c_W)^4}|D_Z|^2\left[(g_L^2+g_R^2)^2 (1+\cos^2\theta)\right. \nonr \\
&&\left.+2(g_L^2-g_R^2)^2\cos\theta\right] \nonr \\
&&+\frac{1}{2 (s_W c_W)^2} \Re(D_{\gamma \eul}^*D_Z )\left[(g_L+g_R)^2 (1+\cos^2\theta)\right. \nonr \\
&&+\left.\left.2(g_L-g_R)^2\cos\theta\right] \right\}\,,
\eeqa
where $\theta$ is the scattering angle of $\mu^-$, $s$ is the center of mass energy squared,
and $c_W(s_W)$ is the cosine(sine) of the weak mixing angle.
Also, the SM Z-lepton couplings are $g_L=-\frac{1}{2} + s_W^2$ and $g_R = s_W^2$.
We have also introduced the dimensionless gauge boson propagator factors,
\beqa
D_{\gamma \eul}&=&1 + \frac{\rho s }{ s-M_X^2 + i M_X \Gamma_X}\,, \nonumber\\
D_Z&= &\frac{s}{s- M_Z^2 + i M_Z \Gamma_Z}\,,
\eeqa
where $M_X$ the mass  and $\Gamma_X$ the width of $Z_l$. We have combined the photon and
$Z_\eul$ exchange  together  since both have vector couplings to $e$ and $\mu$.
The finite widths are included to take care of the behaviors near the mass poles.
The SM $\gamma-Z$ interference causes a wiggling with magnitude around $\sim {\cal O}(10^{-2})$ of the cross section around Z-pole; see Fig.\ref{fig:Z_LS}(a,b). This along with other asymmetries have been
experimentally confirmed by analyzing the Z-line shape\cite{LEP1a,*LEP1b,LEP}.
The presence of the new $Z_\eul$ boson provides additional wiggling around the SM Z-pole at the level of $\sim {\cal O}(10^{-3})$,
Fig.\ref{fig:Z_LS}(b).
This will be the first unambiguous sign of the existing of a new gauge boson which interferes with $\gamma, Z$, which
can be searched for at the future Z factories such as the FCC-ee\cite{FCCEE, *FCCEE_case} and CEPC\cite{CEPC}.

We parameterize the cross section as
\beq
\frac{d\sigma}{d\cos {\theta}}=\frac{\pi \alpha^2}{2s}\left[ A(1+\cos^2{\theta})+B\cos{\theta}\right]\,.
\eeq
Then  $A_{FB}$ is given by
\beq
\label{eq:AFB}
\begin{split}
A_{FB}&=\frac{\int^1_{0} d\cos{\theta} \frac{d\sigma}{d\cos{\theta}}-\int^0_{-1} d\cos{\theta} \frac{d\sigma}{d\cos{\theta}}}
{\int^1_{-1}d\cos{\theta} \frac{d\sigma}{d\cos{\theta}}}\\
&= \frac{3B}{8A}\,,
\end{split}
\eeq
where $A,B$ can be easily read from Eq.(\ref{eq:eeff}). $A_{FB}$ is center-of-mass energy dependent.

At the Z pole, we have
\beq
A_{FB}= \frac{3}{4}\frac{ (g_R^2-g_L^2)^2}{(g_R^2+g_L^2)^2}
\sim 0.01695
\eeq
by using $s_W^2=0.2311$. It is accidentally small because $s_W^2$ is very close to $1/4$.
It is interesting to note that
the $Z_\eul$ exchange induces a universal positive contribution to all $A_{FB}^l$ for SM charged leptons at the Z-pole.
This can be understood because as follows. First, as $Z_\eul$ is heavier than $M_Z$, it gives a destructive interference to the symmetric $D_{\gamma l}$ term and reduces $A$.
 Second, the asymmetric $(g_L-g_R)^2 (\gg (g_L^2-g_R^2)^2)$ term from $Z_l-$SM interference increases $B$ in Eq.(\ref{eq:AFB}).
On the other hand, the $A_{FB}^q$ for SM quarks receive no such contributions. Therefore, with the presence of $Z_\eul$ and $M_X>M_Z$,  $\widetilde{A}_{FB}^l > \widetilde{A}_{FB}^q$, where $\widetilde{A}_{FB}\equiv A_{FB}/A_{FB}^{SM}$, is a robust prediction.
For $A_{FB}^b$, the LEP experimental value is measured to be $0.0992(16)$, and the SM expectation is $0.1031(3)$ \cite{PDG} by using the above value of $s_W$ and all other
SM parameters input from the global electroweak precision fit. Or roughly speaking,
$\widetilde{A}_{FB}^l/\widetilde{A}_{FB}^b = 1.0393 \pm 0.0164$.

However, taking into account the mass bound, Eq.(\ref{eq:MXbound}),  only
a $10.7$ TeV $Z_\eul$ with $\rho\sim 36.3$ can explain this difference between lepton and quark sector at  $2 \sigma$ level.
In other words,  with $g_\eul \sim 6 e$ one can account for this discrepancy.
 Certainly, whether this difference is due to $Z_\ell$ will be clarified
at a future Z-factory option $e^+ e^-$ colliders.

\begin{figure}
    \centering
    \includegraphics[width=0.34\textwidth]{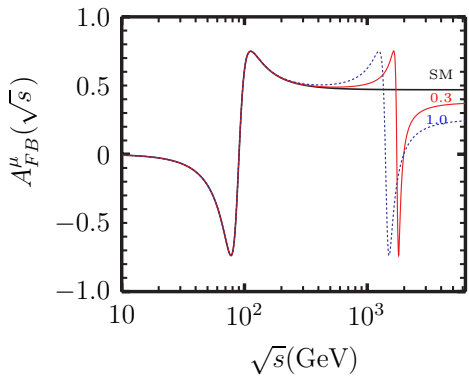}
    \caption{The $A_{FB}$ vs $\sqrt{s}$ with $M_X=2$TeV.
    The black curve is for the SM, and the red(blue) curve is for
     a $Z_\eul$ with $\rho=0.3(1.0)$. }
    \label{fig:AFB}
\end{figure}

Beyond the Z-pole and for $ M_Z \ll \sqrt{s} \ll M_X$,
\beq
\begin{split}
A_{FB} &\sim \frac{3}{4}\, \frac{ (g_R^2-g_L^2)^2 + 2(s_W c_W)^2(g_R-g_L)^2 }{ 4(s_W c_W)^4 +(g_R^2+g_L^2)^2 + 2(s_W c_W)^2(g_R+g_L)^2 } \\
&\sim 0.4691\,.
\end{split}
\eeq
At the $Z_\eul$ pole, the asymmetry is small due to the vector coupling nature of $Z_\eul$ and it becomes
\beq
\begin{split}
A_{FB} &\sim \frac{3}{4}\, \frac{(g_R-g_L)^2}{(g_R+g_L)^2 +2 (s_W c_W)^2 \frac{2}{3\alpha}}\\
&\sim 8.83\times 10^{-6}
\end{split}
\eeq
by using Eq.(\ref{eq:Zlwidth}).
It is interesting that the above three values are not sensitive to $\rho$.
Finally,  for $\sqrt{s} \gg M_X$
\beqa
&&\frac{4}{3}A_{FB} \sim  \\
&&\frac{ (g_R^2-g_L^2)^2 + 2\bar{\rho}(s_W c_W)^2(g_R-g_L)^2}{ 4(s_W c_W)^4\bar{\rho}^2 +(g_R^2+g_L^2)^2 + 2\bar{\rho}(s_W c_W)^2(g_R+g_L)^2 }\,,\nonr
\eeqa
where $\bar{\rho}=1+\rho$.
We give a plot of $A_{FB}$, Fig.\ref{fig:AFB}, for $M_X=2$ TeV and $\rho=0.3, 1.0$.

The $\sqrt{s}$ dependent $A_{FB}$ provides an important handle to probe the new heavy gauge boson.
It is especially useful for the planned linear colliders. For example, the  CLIC has plans for  2-staged intermediate energy at $\sqrt{s}= 0.5, 1.4(1.5)$ TeV before reaching its ultimate $3$TeV goal\cite{CLIC}.
At each new stage, it is able to tune down the energy by about factor 3 without losing the luminosity.
One might be able to see the effect of new gauge boson in the $A_{FB}$ even if $Z_\ell$ is too heavy
to be produced on shell.

\subsection{Muon $g-2$}
The $Z_\eul$ contribution to the muon anomalous moment can be easily calculated to be
\beq
\label{eq:g-2}
\Del a_\mu = \frac{\alpha}{3\pi} \rho \left(\frac{m_\mu^2}{M_X^2}\right)\,.
\eeq
Using the value $a_\mu^{exp}-a_\mu^{th}=2.88\times 10^{-9}$\cite{PDG} and requiring that $\Del a_\mu$ is smaller than that, we obtain the constraint
$M_X>54.5 \sqrt{\rho} $ GeV. The helicity flip factor severely curtails the sensitivity of $a_\mu$ to $M_X$.
 The new exotic scalars have contributions to $\Del a_\mu$ as well. Since their masses have to be heavier than $\sim 0.8$TeV, those contributions are negligible. However, this limit can not compete with Eq.(\ref{eq:MXbound}). For a recent review of the connection between  $a_\mu$ and the new physics, see\cite{Lindner:2016bgg}.

\subsection{M\o eller Scattering}
The exchange of $Z_\eul$ will interfere with the SM $Z,\gm$ processes at the amplitude level.
The leading order is free of hadronic uncertainties and hence offers a very clean sensitive
probe of $Z_\eul$.
Since the $Z_\eul$ admits vector coupling to the electron, it does not contribute left-right asymmetry directly.
Its role in the  M\o eller scattering is to increase  the symmetric cross section from the photon exchange diagram.
 The asymmetry is then reduced to
\beqa
\label{eq:ALR}
A_{LR} &\simeq& A_{LR}^{SM}\times\left[ 1- 6 \frac{\rho  Q^2}{M_X^2}\frac{(1-y)(1-y+y^2)}{ 1+y^4+(1-y)^4} \right]\,,\\
A_{LR}^{SM}&=&\frac{4G_\mu s}{\sqrt{2}\pi\alpha}\frac{y(1-y)}{1+y^4+(1-y)^4}\left[\frac{1}{4}-s_w^2 \right]\,,
\eeqa
where $y=-\frac{t}{s}$.
The asymmetry was measured to be
\beq
A_{LR}= 131\pm 14(\mbox{stas}) \pm 10 (\mbox{syst}) \mbox{ppb}
\eeq
  by the SLAC E158 experiment\cite{SLAC_E158}, where $Q^2=0.026 (GeV)^2$, $y=0.6$, thus $A_{LR}^{SM}=1.47\times 10^{-7}$.
By taking the 95\%C.L. limit, $ -0.337<\frac{\delta A_{LR}}{A_{LR}} < 0.122$.
 Due to the stringent limits from Eq.(\ref{eq:MXbound}),  the $Z_\eul$ contribution to $\frac{\delta A_{LR}}{A_{LR}}$ has no significance at all.

\subsection{$Z_\eul$ production at the LHC}

\begin{figure}[htb!]
\centering
\includegraphics[width=0.4\textwidth]{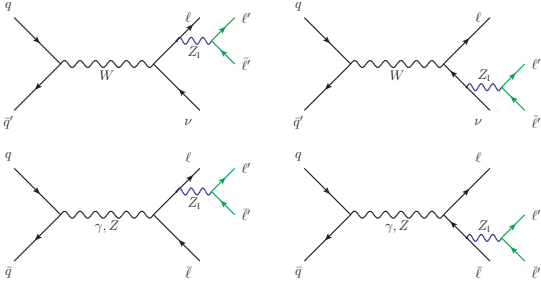}
\caption{Drell-Yan production of $Z_\eul$.}
\label{fig:DYZ}
\end{figure}

If energetically allowed, $Z_\eul$ can be produced at the LHC via the radiaitive Drell-Yan process as depicted
in Fig.(\ref{fig:DYZ}).

The final states will be two pairs of leptons with different flavors in which one pair constitutes
a resonance. E.g. $\mu^+\mu^-$ and $e^+ e^-$ pairs and either pair coming from on-shell
$Z_\eul$ decay. Another signal will be three leptons plus missing energy . A spectacular
example will be a $\mu^+\mu^-$ and an $e^+ e^-$ pair with either pair resulting from $Z_\eul$ decay.
Neither signatures will have no jet activities.

Here, just for illustration purpose, we consider the signal $p p\ra e^+ e^- Z_\eul\ra e^+ e^- +(\mu \bar{\mu})$, and there is a sharp resonance peak of muon pair invariant mass
at $M_X$. The SM background is $p p \ra e^+ e^- \mu^+ \mu^-$ with the $m_{\mu\bar{\mu}}\sim M_X$.
We evaluate the cross section of $p p\ra e^+ e^- Z_\eul$ at LHC for three CM energies by the program CalcHep\cite{CalcHep} with the CTEQ6l1 PDF set\cite{CTEQ}.
The SMBG are also evaluated by CalcHep with a cut that $m_{\mu\bar{\mu}}\in (M_X-50GeV, M_X +50GeV)$.
The numbers are listed in Tab.\ref{tab:DYLHC}.
{\footnotesize {
\begin{table}[ht!]
\centering
\renewcommand{\arraystretch}{2.0}
\begin{tabular}{|c|c|cccc|}
  \hline
$\frac{\sqrt{s}}{\mtev}$  &  & $\frac{M_{X}}{\mtev}=0.5$ & $\frac{M_{X}}{\mtev}=1.0$ &$\frac{M_{X}}{\mtev}=2.0$ &$\frac{M_{X}}{\mtev}=5.0$  \\
   \hline
14& $\frac{\sigma}{g_l^2}$ &  $ 5.4 \times 10^{-5}$ & $1.7 \times 10^{-6}$ &$1.9 \times 10^{-8}$ &$9.8 \times 10^{-13}$ \\
&$\sigma_{BG}$ &  $ 2.2 \times 10^{-5}$ & $1.4 \times 10^{-6}$ &$5.4 \times 10^{-8}$ &$6.2 \times 10^{-11}$ \\
&$\frac{\bar{w}_{max}}{\mtev}$ &  $0.61$ & $0.43$ & - & -\\
  \hline
30& $\frac{\sigma}{g_l^2}$ &  $ 2.6 \times 10^{-4}$ & $1.5 \times 10^{-5}$ &$5.1 \times 10^{-7}$ &$1.0 \times 10^{-9}$ \\
&$\sigma_{BG}$ &  $ 6.8 \times 10^{-5}$ & $7.1 \times 10^{-6}$ &$4.2 \times 10^{-7}$ &$5.7\times 10^{-9}$ \\
&$\frac{\bar{w}_{max}}{\mtev}$ &  $ 1.02$ & $0.85$ & - & -\\
  \hline
100& $\frac{\sigma}{g_l^2}$ &  $ 1.7\times 10^{-3}$ & $1.5\times 10^{-4}$ &$1.1\times 10^{-5}$ &$1.8 \times 10^{-7}$ \\
&$\sigma_{BG}$ &  $ 3.0\times 10^{-4}$ & $3.2\times 10^{-5}$ &$2.8\times 10^{-6}$ &$7.6\times 10^{-8}$ \\
&$\frac{\bar{w}_{max}}{\mtev}$ &  $ 1.79$ & $1.85$ &$1.83$ & -\\
  \hline
\end{tabular}
  \caption{The $pp\ra e^+e^- Z_l$ cross section normalized by $g_l^2$ and the SM BG.
  The cross sections are in $(pb)$, and $\bar{w}_{max}$ are in TeV.}
\label{tab:DYLHC}
\end{table} }}

We use $S/\sqrt{B}=3$ and an integrated luminosity ${\cal L}_0=3000 (fb)^{-1}$ as the benchmark limit of detecting a $Z_l$ at the LHC.
Then,
\beq
\begin{split}
3&= \frac{\sigma(pp\ra e^+e^-Z_\eul(\mu\bar{\mu}))  Br(Z_\eul\ra \mu\bar{\mu})\times {\cal L}_0}
 {\sqrt{ \sigma_{BG} \times {\cal L}_0 } }\\
& = \frac{2}{9} \sqrt{\frac{{\cal L}_0}{ \sigma_{BG}} } \, \left(\frac{\sigma}{g_\eul^2}\right)\, g_\eul^2\,.
\end{split}
\eeq

The corresponding highest lepton number breaking scale we can probe is
 $\bar{w}^2_{max} = (2/27) \sqrt{ {\cal L}_0 /\sigma_{BG} }\times(\sigma/g_\eul^2) M_X^2$, which is also displayed in Tab.\ref{tab:DYLHC}.
The LHC14, LHC30, and LHC100 have the potential to probe the lepton-number violating scales up to
$\sim 0.5, 1.0$, and $2.0$ TeV, respectively.

\section{ Radiative Seesaw mass for $\nu_L$}
\begin{figure}[htb!]
 \centering
 {\includegraphics[width=0.4\textwidth]{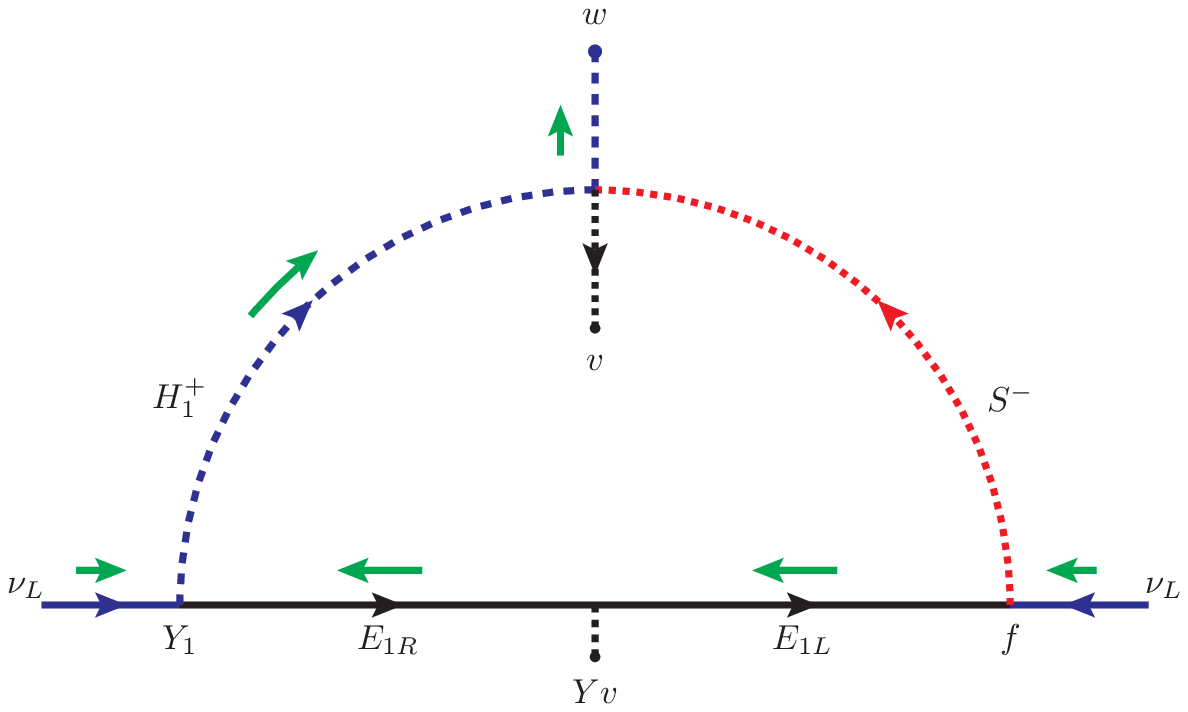}}
 \caption{one-loop $\nu_e \, \nu_e$ mass term generation. Green arrows show the flow of lepton charge}
 \label{fig:nunumass}
 \end{figure}

The Feynman diagrams for radiative $\nu_L$ mass generation are depicted in Fig.\ref{fig:nunumass} which are
 given in the weak eigenbasis. They fill in the upper left-hand block of zeros in Eq.(\ref{eq:numm})
 \footnote{Our anomaly solutions can accommodate a variant of
the type-I seesaw mechanism by adding a set of vectorlike SM singlet neutrinos $\mathcal{N}_R$ and $\mathcal{N}_L$ with lepton number unity. We shall not pursue this further.}.
 \begin{figure}[htb!]
 {\includegraphics[width=0.4\textwidth]{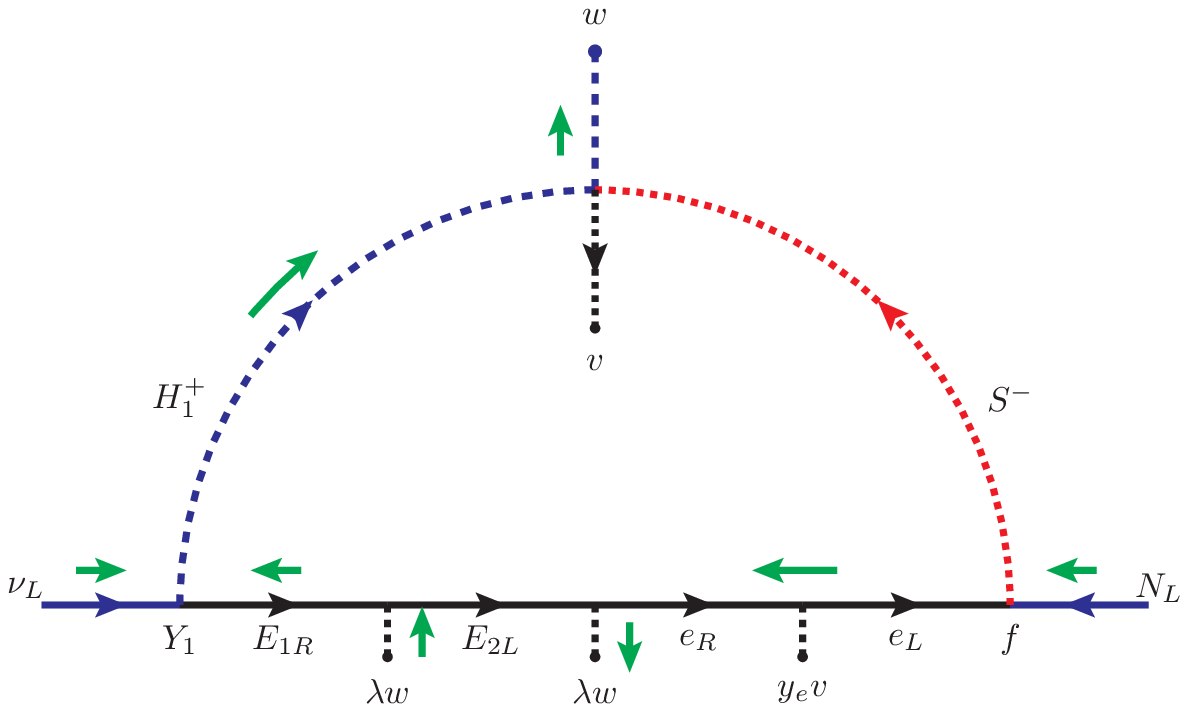}}
\caption{one-loop $\nu_{eL}\,N$ mass term generations. Green arrows show the flow of lepton charge.}
\label{fig:nuNmassnN}
 \end{figure}

 In the limit that the charged $H_{1},S$ scalars are heavier than the leptons, we get
 \beq
 M_{11}=\frac{f (Y_1 v)(\lam_{1\ell} w)m_e}{16\pi^2(m_1^2-m_S^2)}\ln{\left(\frac{m_1^2}{m_S^2}\right)}\,,
 \eeq
where we have used $Y v=m_e$ as explained above. $M_{11}$ will be the upper leftmost entry
in Eq.(\ref{eq:numm}).
Similarly, we get radiative correction to $M_{12, 21}$, Fig.\ref{fig:nuNmassnN}.
Clearly, these are much smaller than $\bar{\lam} w$. Other than providing a Majorana mass for
the active neutrino $\nu_e$, they also transform the Dirac neutrino $N$ into a pseudo-Dirac one.
Numerically, the splitting will be undetectably small for all practical purposes, and we can treat
the $N$ as a Dirac neutrino.
Assuming that there is no outstanding hierarchy between $m_1$ and $m_S$, then one expects the combination
$w/(m_1^2-m_S^2) \ln(m_1^1/m_S^2) \simeq w \sim {\cal O}(\mbox{TeV})$. Plugging in the values, the resulting active neutrino mass
is around $m_\nu\sim f Y_1 \lambda_{1\ell} \times 10^3$ eV. And the sub-eV neutrino mass can be easily achieved with $f,Y_1,\lambda_{1\ell} \sim {\cal O}(0.1)$
without prominent fine-tuning.
\section{Phenomenology of $E,N$}

Even if the exotic leptons are too heavy
to be produced by current or near-future
colliders, they can have important effects
at current energies. The notable ones are
the electroweak oblique parameters $S,T$\cite{S_T_1, S_T_2}, and the decay $h\to \gm\gm$.

\subsection{ Oblique parameters $S,T$}
\label{section:S_T}
It is well known that the oblique parameters $S$ and  $T$ constraint heavy
fermions that carry SM quantum numbers. In this case, they constrain the mass differences
of the lepton pairs $E,N$ as well as the number of such pairs.
Explicitly, for each generation we have
\beqa
\label{eq:ObT}
\triangle T&=&\frac{1}{16\pi s_w^2 }\sum_{i=1,2}\frac{M_{E_i}^2}{M_W^2}\left(1+x_i+\frac{2x_i}{1-x_i}\ln x_i\right)\,,\\
\label{eq:ObS}
\triangle S&=& \frac{1}{6\pi}\sum_{i=1,2}\left(1+ \ln x_i\right)\,,
\eeqa
where $x_i = M_{N_i}^2/M_{E_i}^2$.
When the mass splitting between $E_i$ and $N_i$ is small comparing to their masses,
\beqa
\label{eq:OBTL}
\triangle T &\sim & \frac{1}{12\pi s_w^2 M_W^2}[(M_{N_1}-M_{E_1})^2+(M_{N_2}-M_{E_2})^2 ]\,, \nonr\\
\triangle S &\sim & \frac{1}{3\pi} \left[1+\frac{M_{N_1}-M_{E_1}}{M_{E_1}}+\frac{M_{N_2}-M_{E_2}}{M_{E_2}} \right]\,,
\eeqa
 for each generation.
The doublet $H_1$ provides contribution
\beqa
\label{eq:OBTS}
\triangle T&=&\frac{1}{16\pi s_w^2} \frac{M_{H^+}^2}{M_W^2}\frac{1}{z}  \left[ 1+ z +\frac{2z}{1-z}\ln z \right]\,,\\
\label{eq:OBSS}
\triangle S&=& -\frac{1}{12\pi} \ln z\,,
\eeqa
where $z \equiv M_{H^+}^2/M_{H_1^0}^2$. Note that  $\triangle T$ from  fermions and doublet scalar are both positive,
but $\triangle S$ from the doublet scalar can be either positive or negative.
 From the Particle Data Group, we have $S_{data}<0.22$ and $T_{data}<0.27$ at 95\% C.L.\cite{PDG}

\begin{figure}[htb!]
 \centering
 \includegraphics[width=0.4\textwidth]{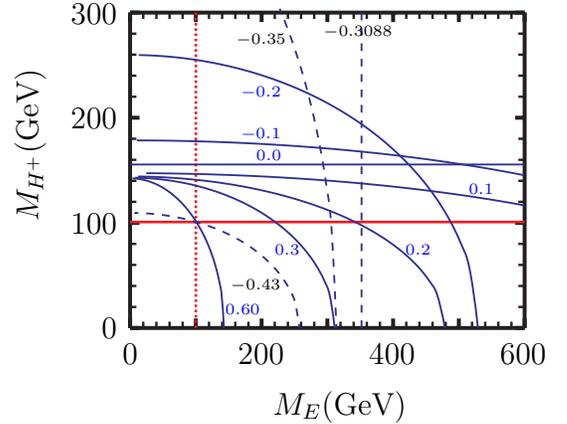}
 \caption{Contours from $\triangle S$ and $\triangle T$ constraints for different $\ln x$.
  For a given $\ln x$, the allowed masses region is above the direct search bound on $M_{H^+}$, the horizontal red line,
   to the right of the direct search bound on $M_E$, the vertical dashed red line, and to the lower left of the blue (dash)curve.}
 \label{fig:ST_contour}
 \end{figure}

 To see how these will restrict the parameters
 of our model, we begin by taking $z=1$; i.e. the neutral and charged components of $H_1$ are degenerate. This implies that the mixing of $H_1$ with all other scalars is
 negligible. Then, the scalar contributions to $\triangle T $ and $\triangle S$ are vanishing. For simplicity, we also assume  the masses of $E_+$ and $E_{-}$ are equal and their counterparts for $\mu$ and $\tau$ families are also the same. From Eq.(\ref{eq:ObS}) we see that the new isodoublet chiral leptons cannot have degenerate upper and lower components; otherwise it runs afoul of $S_{data}$. The splitting between the neutral and charged components that
 saturates $T_{data}$ is given by
 \beq
 x = 0.73
 \eeq
 where we have dropped the subscript $i$.
Using  Eq.(\ref{eq:ObT}) and $T_{data}$, we obtain $M_E \leq 350 $ GeV. The above
values are to be taking as a demonstration that the stringent constrains of oblique corrections
can be satisfied with new lepton masses in the range of 350 GeV,  the vertical dash blue line in Fig.\ref{fig:ST_contour}. This is well above limit form the charged lepton searches of $> 100.8$ GeV,  the vertical dash red line in Fig.\ref{fig:ST_contour}, given in PDG\cite{PDG}
\footnote{ Note that the limit, $M_E\gtrsim 168$GeV at $95\%$ C.L., obtained by \cite{ATLAS_heavy_L} does not apply here since there is no tree-level  $Z\mhyphen E \mhyphen l$ coupling  in our model. }.

We can  also of take the limit that the  $E$ and $N$ are degenerate, i.e.
all $x=1$. Then Eq.(\ref{eq:ObS}) yields  $\triangle S \simeq .32$. Then, it will require
$H_1$ doublet with  $\ln z \simeq 3.71$ to bring it within the experimental bound since scalars give a negative contribution (see Eq.(\ref{eq:OBSS})). The scalars will then be the sole contributors to $\triangle T$. A similar calculation gives an upper bound on the mass of $H^+$ to be 155 GeV,  the horizontal blue line in Fig.\ref{fig:ST_contour}.
This value is also larger than the direct search bound on charged scalars,  the horizontal red line in Fig.\ref{fig:ST_contour}. The large splitting
between $H^+$ and $H_1^{0}$ implies that some if not all of the parameters $\kappa_{H_1 S},\kappa_{\Phi_i H_1}$ and $\mu_3$ in Eq.(\ref{eq:VA})are large.

Notice that since $z$ is large, we have $M_{H_1^{0}}\simeq 24.4$ GeV. In general, it can mix with the SM Higgs boson, but we have seen that this mixing is constrained  to be small.  In the interaction basis, which is good for small mixings, it does not couple to quarks, and it has no couplings
to gluons to one-loop. Because of this, it will not run into problem at the LHC.
 Neutral scalars in the mass range of $10 -100$ GeV is notoriously difficult to detect. The challenges and a possible
signal for probing this at the LHC was discussed in \cite{CMN}. Additionally,  a more
promising avenue of exploring this at a future $e^+ e^-$ collider is given in \cite{CNW}.

From the above limiting case, it is easy to see that if $\ln x > -0.3088$ the doublet $H_1$ will
have to play a role in satisfying the oblique corrections bounds. For $\ln x=-0.3088$, it is the vertical dashed blue line in Fig.\ref{fig:ST_contour}.
It is more realistic to assume finite mass splittings between the isodoublets for both leptons and scalars. As an illustration we take  $\ln x= -0.2$, then
the masses of $M_E$ and $M_{H^+}$ will satisfy a contour given by
\beq
 .0062\left(\frac{M_E}{M_W}\right)^2 + .0259\left(\frac{M_{H^+}}{M_W}\right)^2 \leq 0.27.
\eeq

The allowed regions of $M_E$ and $M_{H^+}$ for different $\ln x$ are displayed in Fig.\ref{fig:ST_contour}.
If both experimental lower bounds on $M_E, M_{H^+}$ are met, one can see that  the range of $\ln x$ is $-0.43572<\ln x <0.6053$, and it implies that
 $ 0.80 < (M_N/M_E) < 1.35 $ and $ 0.47 <(M_{H^+}/M_{H^0})<241.0$ if a universal $x_i$ is assumed.
Moreover, for  $\ln x > -0.1$, a light neutral scalar with mass $M_{H^0}<50$ GeV is expected.

 Before closing this section, we remark that it is well-known that the constraint from $\triangle T $ can be largely loosen by introducing an $SU(2)$ triplet with a small VEV so that the tree-level electroweak $\rho$-parameter is less than unity. However, for the minimal setup, we do not further into such discussion. See \cite{Chang:2018wsw} for utilizing a triplet Higgs for neutrino mass generation and relaxing the $\Delta T$ constraint in this model.

\subsection{Impact on Higgs decays}
\subsubsection{ Higgs to two photons.}
The SM Higgs to the di-photon vertex is generated at the one-loop level with dominant contributions from $W^\pm$ and top quark running in the loop.
In our model, there are six new charged leptons; a $E_1, E_2$ pair for each of the three generations, and two new charged scalars, $H_1^\pm$ and $S^\pm$.
These charged fields mix among themselves, and one needs to know every parameter for the actual mass diagonalization.
However, with assumptions on the masse ranges of these charged particles,  a general discussion is sufficient to draw qualitative conclusions.

In the mass basis, we  can parameterize the Yukawa couplings and cubic couplings to the SM Higgs as
\beq
{\cal L}\supset - \sum_{i=1}^6 y_{E_i} \bar{E}_i E_i h - \sum_{i=1,2} \lambda_i M_W h H_i^+H_i^-\,.
\eeq
These new electrically charged degrees of freedom enter the one-loop triangle diagram and modify the width of the SM Higgs di-photon decay.
This is given by \cite{tome}
\beqa
\Gamma(H\ra \gamma\gamma)= {G_F \alpha^2 M_H^3 \over 128\sqrt{2} \pi^3}
\left| F_1(\tau_W) +\frac{4}{3}F_{1/2}(\tau_t)\right.\nonr\\
+\sum_{j=1,2} \lambda_i \frac{M_W^2}{g_2 M_{H_i}^2}F_0(\tau_{H_i})\nonr\\
\left.+\sum_{i=1}^6  y_{E_i}\frac{2 M_W}{g_2 M_{E_i}}F_{1/2}(\tau_{E_i})
\right|^2\,,
\eeqa
where $\tau_i\equiv (m_H/2 m_i)^2$, and
\beqa
F_0(\tau)&=&-[\tau-f(\tau)]/\tau^2\,,\nonr\\
F_{1/2}(\tau)&=&2[\tau+(\tau-1)f(\tau)]/\tau^2\,,\nonr\\
F_1(\tau)&=&-[2\tau^2+3\tau+3(2\tau-1)f(\tau)]/\tau^2\,,
\eeqa
with
\beq
f(\tau)= \left\{ \begin{array}{ll}
                   \left[\sin^{-1}\sqrt{\tau}\right]^2\,, & \mbox{if}\, \tau\leq 1\\
                   -\frac{1}{4}\left[\log{1+\sqrt{1-1/\tau} \over 1-\sqrt{1-1/\tau}}-i\pi\right]^2\,, & \mbox{if}\, \tau> 1\,.\\
                 \end{array}
\right.
\eeq
For $m_i\gg M_H/2=62.5$GeV, we have the following expansions around $\tau=0$:
\beqa
F_0(\tau)&\sim & \frac{1}{3}+\frac{8}{45}\tau +{\cal O}(\tau^2)\,,\nonr\\
F_{1/2}(\tau)&\sim & \frac{4}{3}+\frac{14}{45}\tau+{\cal O}(\tau^2)\,,\nonr\\
F_1(\tau)&\sim & -7-\frac{22}{15}\tau+{\cal O}(\tau^2)\,.
\eeqa
 Plugging in the numbers,
the di-photon decay width reads
\beqa
\Gamma(H\ra \gamma\gamma)= \left.{G_F \alpha^2 M_H^3 \over 128\sqrt{2} \pi^3}
\times \right| -8.324 + 1.834 \nonr\\
+    8.3\times 10^{-4}(1.3 \times 10^{-2} )\times \lambda_2 +  0.087(0.42) \times  \lambda_1  \nonr\\
\left.+\sum_{i=1}^6   0.32(3.64 )\times y_{E_i}  \right|^2
\eeqa
 for  $M_{E_i}=1000(100)$GeV, $M_{H_2}=2.0(0.5)$TeV, and $M_{H_1}=200(100)$GeV.
The first two numbers are the dominate SM contributions from $W^\pm$ and the top quark, respectively.
Since we expect $ |y_{E_i}| \sim m_l/v_h \ll 1$, even  for the new leptons (see Sec. III), the charged leptons contribution can be ignored.
If the second charged scalar is heavy,  its contribution can be ignored, too, even taking $\lambda_2\sim{\cal O}(1)$.
Therefore, only the light charged scalar with mass in the range
of 100 to 200 GeV matters.
The gluon fusion is the dominant SM Higgs production channel at the LHC, and it does not receive any modification. The signal strength of $pp\ra h\ra \gamma\gamma$ at the LHC is therefore
\beqa
\mu_{\gamma\gamma}&\simeq &\Gamma(H\ra \gamma\gamma)/\Gamma(H\ra \gamma\gamma)_{SM}\nonr\\
 &\sim& 1 - (0.03-0.13)\times \lambda_1\,.
\eeqa
Comparing with the experiment data $\mu_{\gamma\gamma}=1.18 (+0.17 -0.14)$ \cite {CMS18}, we conclude that it is safe even the light charged Higgs has a coupling $|\lambda_1|\sim{\cal O}(1)$. This is in agreement with the general analysis given in \cite{CNWS}.
\subsubsection{ Higgs to 4 fermions}
For notational simplicity, here we denote $h_1\equiv \Re(H_1^0) $ and $a_1 \equiv \Im(H_1^0)$.
If $h_1$ ( or $a_1$ ) is lighter than half the mass of the SM Higgs, $h_{SM}$, then we can have
\beq
h_{SM}\ra 2 h_1 ( 2 a_1 ) \ra \bar{\ell_i}\ell_i +\bar{\ell_j}\ell_j\,.
\eeq
The decay width is
\beq
\Gamma(h_{SM}\ra h_1 h_1(a_1 a_1))=\frac{v^2(\kappa_2+\kappa_3)^2}{32\pi M_H}\left(1-\frac{4 m_1^2}{M_H^2}\right)^{\frac{1}{2}}\,,
\eeq
where $M_H(=125\mbox{GeV})$ and $m_1$ are the masses of $h_{SM}$ and $h_1(a_1)$, respectively. We have neglected term involving off-diagonal mixing of neutral scalars.
The dominant decay mode for $h_1(a_1)$ is model dependent.
The current bound on the mixing squared between $h_{SM}$ and $h_1$ is about $\lesssim 10^{-2}$ for $10 <m_1<40$ GeV from LEP2\cite{LEP2_higgs_search}\footnote{The limit on the mixing squared between $h_{SM}$ and $h_1$ could be improved by a few orders of magnitude at the future colliders\cite{CMN,CNW}. }.
For $Y_1\sim O(0.1)$,  as expected from the radiative neutrino masses, the effects from the mixing with SM Higgs  cannot compete with those from the direct Yukawa interaction, Eq.(\ref{eq:H1Yukawa_A}).
Therefore, for scenario A , the main decay channel will be $h_1(a_1)\ra \bar{\ell} \ell$
with $\ell=e,\mu,\tau$.  The signal will be SM Higgs decays into 2 charged leptons pairs  with both invariant masses peaking at the unknown $m_1$.

For scenario B, $h_1(a_1)$ has only off-diagonal couplings to SM lepton and a heavy
lepton; see Eq.(\ref{eq:H1Yukawa_B}).  The dominant decay of light $h_1(a_1)$
is expected to be due to mixing with the $h_{SM} ( \Im(H_0^0))$   which then decays
into a fermion pair.  Therefore, $\bar{b} b$  will be the dominate final state if $m_1 > 10  $ GeV.

For the general case, in between scenario A and B, we expect the mixing element $|U_{12}|^2 < 0.13$, from unitarity and $|U_{11}|^2>0.87$ \cite{LHCmu}, to be small but non-vanishing.

\subsection{Colliders production and decay of exotic leptons}
For both scenarios (A) and (B), the SM gauge interaction
allows  $E_\pm\ra W^- N_\pm, W^- N_\mp$, and $N_+\ra N_- Z$ if $M_N<M_E$ is assumed.
We consider the decay, $E_\pm \ra N W^-$, of a heavy Dirac $N$ for simplicity.   The decay width is calculated to be
\beqa
\Gamma_{E_\pm \ra N W^-}&=&
{G_F M_E^3 \over 8\sqrt{2} \pi}
\lambda_{cm}(x_N,x_w)f(x_w,x_N) \nonr \\
f(x,y)&=&x(1+y-2x \mp 6\sqrt{x}) +(1-y)^2
\eeqa
where $\lambda_{cm}(y,z)=\sqrt{1+y^2+z^2-2(y+z+y z)}$, $x_w=(M_W/M_E)^2$, and $x_N \equiv (M_N/M_E)^2$.
For $M_E \gg M_W$, or $x_w\ll 1$, the width becomes
\beq
\Gamma_{E_\pm \ra N W^-} \simeq {G_F M_E^3 \over 8\sqrt{2} \pi}(1-x_N)^3\,.
\eeq

As discussed in Sec.\ref{section:S_T}, the oblique corrections requires that $\ln x_N>-0.4357$, which implies that
\beq
\Gamma_{E_\pm \ra N W^-} < 14.45 \times \left( {M_E \over 1 \mbox{TeV}}\right)^3 \,\mbox{GeV}\,.
\eeq
On the other hand, if $M_N>M_E$, the decay width of $N$ takes a similar form with $M_E\leftrightarrow M_N$,
\beq
\Gamma_{N\ra E_\pm W^+} \simeq {G_F M_N^3 \over 8\sqrt{2} \pi}(1-x_N^{-1})^3\,.
\eeq
Similarly, from that $\ln x_N< 0.6053$,
\beq
\Gamma_{N\ra E_\pm W^-} < 30.72 \times \left( {M_N \over 1 \mbox{TeV}}\right)^3 \,\mbox{GeV}\,.
\eeq
From the above discussion, unless the leptons are nearly degenerate, the decays of $E_\pm$ or $N$ are expected to be prompt.

Next, we turn our attention to the heavy lepton decay via the Yukawa interaction with $H_1$.
Let's consider a general case with two fermions $F,f$, and a scalar $\phi$. The scalar field $\phi$ could be either neutral or charged.
Assume that they admit a  Yukawa interaction which is parameterized as ${\cal L}\supset \bar{F} (s + a\gamma^5)f \phi +h.c.\,$.
 If kinematics allowed,  the decay  channel $F\ra f \phi$ opens and the width is calculated to be
\beqa
{\lambda_{cm}(x_f,x_\phi) M \over 16\pi}&& \left[(1+x_f)^2 |s|^2\right.\\
&& \left. +(1-x_f)^2 |a|^2 -x_\phi (|s|^2+|a|^2)\right]\,,\nonr
\label{eq:NE_H1decay}
\eeqa
where $M$ is the mass of $F$, $x_f=(m_f/M)^2$, and $x_\phi=(m_\phi/M)^2$.
For the cases that $M,m_f \gg m_\phi$, the decay width becomes
\beq
\Gamma(F\ra f \phi) \simeq \frac{M}{16\pi}(1-x_f)\left[(1+x_f)^2 |s|^2 +(1-x_f)^2 |a|^2 \right]\,.
\eeq
The relevant fields and Yukawa couplings in our model are collected and listed in Tab.\ref{tab:Yukawa_H1}.
One can read the precise expression by using  Eq.(\ref{eq:NE_H1decay}) and Tab.\ref{tab:Yukawa_H1}.
Roughly speaking, the decay widthes are about $\sim (Y_1^2 /64 \pi )M$, or numerically $\sim 0.05 \times(Y_1/0.1)^2\times(M/1\mbox{TeV})$ GeV, where $M$ is the mass of $E_\pm$ or $N$. In general, this decay width is much smaller than that from the decay with a SM $W^\pm$ boson in the final states.
Note that in scenario-B, $N$ does not have the tree-level  two-body decays via the Yukawa interaction with $H_1$\footnote{We have checked that even $M_N<M_E$ $N$ can not be a dark matter candidate due to its SM $SU(2)$ interaction.  Adding an ad hoc $Z_2$ parity will not change this.}.
However, if the mass of the  charged scalar $ S^\pm$ is less than $M_N$, then the decay $N\ra e^+ S^-$ \footnote{ Due to the radiative generated Majorana masses, Fig.\ref{fig:nuNmassnN}, $N$ is in fact pseudo-Dirac. However, the conjugate
decays, $N\ra W^-\bar{E}$ and  $N\ra e^- S^+$, are expected to be rare.} is possible. Otherwise $N$ will have only three-body decays. For $M_N<M_E$, there is another chain with an intermediate virtual $E$; for example, $N\ra W^+ E^* \ra W^+ h_1 e^-$. We shall use the superscript ``$*$'' to denote off-shell particles.
\begin{table}
\begin{center}
\begin{tabular}{|c|ccccc|}
\hline
scenario & $F$ & $f$ & $\phi$ & $s$ & $a$\\
\hline
 & $E_\pm$ & $ e$ & $h_1$ & $0$ & $-\frac{Y_1}{2\sqrt{2}}$\\
(A)  & $E_\pm$ & $ e$ & $a_1$ &  $-\frac{i Y_1}{2\sqrt{2}}$ & $0$\\
  & $E_\pm$ & $ \nu $ & $H_1^-$ &  $\frac{ Y_1}{4\sqrt{2}}$ &  $-\frac{ Y_1}{4\sqrt{2}}$\\
  & $N$ & $ e $ & $H_1^+$ &  $-\frac{ Y_1}{4}$ &  $-\frac{ Y_1}{4}$\\
\hline
 & $E_\pm$ & $ e$ & $h_1$ & $\frac{Y_1}{4}$ & $-\frac{Y_1}{4 }$\\
(B)  & $E_\pm$ & $ e$ & $a_1$ &  $-\frac{i Y_1}{4}$ & $\frac{i Y_1}{4}$\\
  & $E_\pm$ & $ \nu $ & $H_1^-$ &  $\frac{ Y_1}{2\sqrt{2}}$ &  $-\frac{ Y_1}{2\sqrt{2}}$\\
\hline
\end{tabular}
\caption{The $H_1$ Yukawa couplings between the heavy leptons and the SM ones in our model. The heavy neutrino is assumed to be Dirac.}
\label{tab:Yukawa_H1}
\end{center}
\end{table}

 \begin{figure*}[htb!]
 \centering
 \subfigure[]{\includegraphics[width=0.34\textwidth]{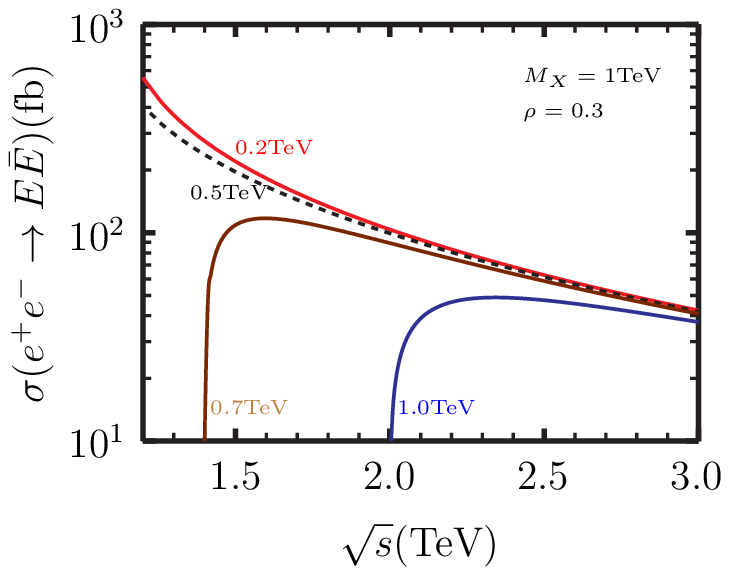}}
  \subfigure[]{\includegraphics[width=0.34\textwidth]{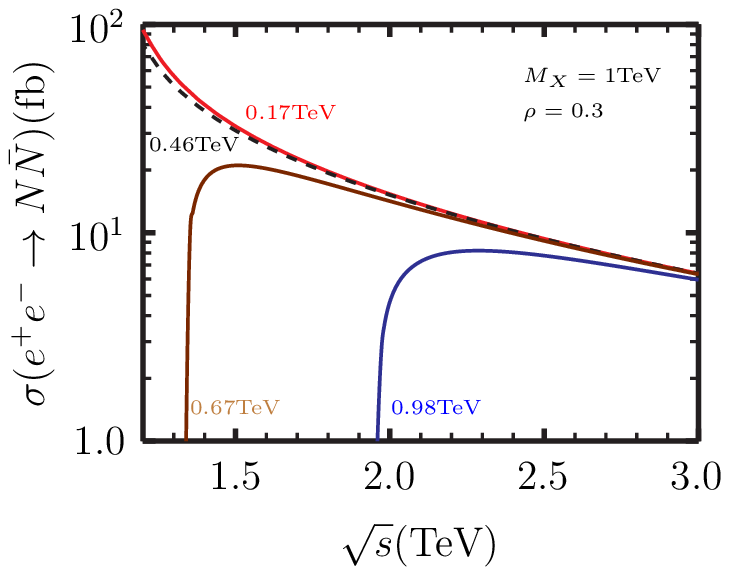}}
\caption{ (a) The $E\bar{E}$, and (b)$N\bar{N}$ production cross sections v.s $\sqrt{s}$ at an $e^+e^-$ collider.
The masses of $E(N)$ are labeled next to the curves.
    }
\label{fig:ee_EN_prod}
 \end{figure*}

The heavy leptons can be pair produced at the $e^+e^-$ colliders. For simplicity, we assume that $E_+$ and $E_-$ are nearly degenerate so that they are hard to be distinguished experimentally, and we collectively denote the states as $E \equiv E_+\sim E_-  $.
For  $\sqrt{s} \gg M_Z$ and away from the $Z_\ell$ pole, the production cross section  per generation for scenario-(B) can be calculated to be
\beqa
&&\sigma(e^+e^-\ra E\bar{E}) \simeq\nonr\\
&&2{4\pi\alpha^2 \over 3s}\sqrt{1-4x_E}\left\{ \left(1+{\rho s\over s-M_X^2}\right)^2 ( 1+2x_E )\right.\nonr\\
&&+{ (g^e_L)^2+(g^e_R)^2\over 4 (s_W c_W)^4 } \left[ ( 1-x_E)((g^E_L)^2+(g^E_R)^2) +6x_E g^E_L g^E_R \right]\nonr\\
&&+\left.{(g^e_L + g^e_R )\over 2 (s_W c_W)^2}(g^E_L + g^E_R )\left(1+{\rho s\over s-M_X^2}\right)\left( 1-x_E\right)\right\}
\eeqa
where $x_E\equiv M_E^2/s$, $g^E_L\simeq g^e_L=-1/2+s_W^2$, $g^E_R\simeq g^e_R =s_W^2$.
The first factor 2 represents the incoherent sum from the contribution of the two heavy Dirac charged leptons.
 For example, if $\{M_{E_+},M_{E_-}\} =\{200,180\}$GeV, $\rho=0.3$, and $M_X=1$TeV, the production cross section of 2 charged leptons are $556.0(83.2)$ fb for
$\sqrt{s}=1.2(2.2)$TeV, see Fig.\ref{fig:ee_EN_prod}(a).
Similarly, the heavy $N$  can be pair produced through the s-channel process mediated by $Z, Z_\ell$, and the  production cross section for each generation is
\beqa
&&\sigma(e^+e^-\ra N\bar{N}) \simeq\nonr\\
&&{4\pi\alpha^2 \over 3s}\sqrt{1-4x_N}\left\{ \left({\rho s\over s-M_X^2}\right)^2 ( 1+2x_N )\right.\nonr\\
&&+{ (g^e_L)^2+(g^e_R)^2\over 8 (s_W c_W)^4 } ( 1+2 x_N) \nonr\\
&&+\left.{(g^e_L + g^e_R )\over 2 (s_W c_W)^2} \left({\rho s\over s-M_X^2}\right)\left( 1-x_N\right)\right\}
\eeqa
 For example, if $M_N = 170$GeV, $\rho=0.3$, and $M_X=1$TeV, the production cross section of $N\bar{N}$ pair are $94.0(12.4)$ fb for
$\sqrt{s}=1.2(2.2)$TeV, see Fig.\ref{fig:ee_EN_prod}(b).

At the LHC, the heavy leptons can be produced via the  photon and(or) W-/Z-boson Drell-Yan process. Therefore, the production cross sections are independent of the $Z_\ell$ mass and the $U(1)_\ell$ gauge couplings.
Since a full-fledged collider study is beyond the scope of this paper, only the production cross-section are considered here.
Three sets of $(M_{E_+}, M_{E_-})$ are considered as the benchmarks with the assumption that their mass differences are sub-electroweak.
For each benchmark, $M_N$ is chosen to meet the constraint from the electroweak precision;
with  $|\ln x| < \{ 0.3, 0.15, 0.05\}$  for $M_{E_+}=\{200,500,1000\}$ GeV, respectively.
The production cross sections are evaluated by CalcHep and listed in Table\ref{tab:LHC_EE_production} and Table\ref{tab:LHC_EN_production}.

 For $M_E, M_N \lesssim 500$ GeV, the production cross sections are about ${\cal O}(1-100) fb$.
The production of $E$ and $N$ will be followed by their decays into SM particles. The decay modes are
sensitive to the masses of $E_+, E_{-}, N$ as well as the masses of the charged scalars $H_1^{\pm}, S^\pm$, which in general mix. If the splitting between $E,N$ is large enough the dominant decays will be two-body modes; otherwise, they will be three-body modes. They also depend on the ordering of the $E,N$ masses.
 If $M_E > M_{S^\pm} $,  then we have the chain
 \beqa
E\ra \bar{\nu} +&S^-& \nonr \\
&\rotatebox[origin=c]{180}{\Large{$\Lsh$}}& W^- +h_1(a_1)\,,
\eeqa
where the decay of $S^-$ proceeds via mixing with $H_1^-$.
Additionally, in scenario-(B), if $M_E > M_N$  we also have the chain
\beqa
E\ra W^- +& N & \nonr \\
        &\downarrow & \nonr \\
\phantom{E\ra} W^-  +     & E^{(*)} & + W^+ \nonr \\
&\rotatebox[origin=c]{180}{\Large{$\Lsh$}} & e^- + h_1(a_1)\,,
\eeqa
where we take
the decay of $E$ to proceed via
\beq
E\ra e+h_1(a_1)\ra e + \ell \bar{\ell}\,
\eeq
if $Y_1$ is not too small.

Next, we examine the decays of the neutral lepton $N$. If $M_N>M_E$, the decay chain will be
\beqa
\label{eq:NWE}
N\ra W^+ + & E & \nonr \\
      &\;\rotatebox[origin=c]{180}{\Large{$\Lsh$}}&
  e^- +h_1(a_1) \,.
\eeqa

Similar to the case of $E$, if $M_N >M_{S^\pm}$  the following is also available
\beqa
\label{eq:NSe}
N\ra e^+ +&S^-& \nonr \\
&\;\rotatebox[origin=c]{180}{\Large{$\Lsh$}} & W^- h_1(a_1) \, .
\eeqa

 Before one can draw any conclusion, it is crucially important to understand the SM background first. We leave the comprehensive signal and background study to a future work.
\begin{table}
\begin{center}
\begin{tabular}{|c|cc|ccc|}
\hline
& $M_{E_+}$ & $M_{E_-}$ & $\bar{E}_+ E_+ $ & $\bar{E}_+E_- +\bar{E}_- E_+$ & $\bar{E}_- E_-$ \\
\hline
set-1 & $200$ & $180$ & $9.62\times 10^{-2}$ & $4.27\times 10^{-2}$ & $1.40\times 10^{-1}$\\
set-2 &$500$ & $480$ & $2.76\times 10^{-3}$ & $9.04\times 10^{-4}$ & $3.30\times 10^{-3}$\\
set-3 &$1000$ & $950$ &$9.08\times 10^{-5}$ & $2.44\times 10^{-5}$ & $1.21\times 10^{-4}$\\
\hline
\end{tabular}
\caption{The production cross sections(in pb) at the LHC14 for one generation in scenario-B in our model. The masses of $E_\pm$ are in the unit of GeV. }
\label{tab:LHC_EE_production}
\end{center}
\end{table}

\begin{table}
\begin{center}
\begin{tabular}{|c|c|ccc|}
\hline
& $M_N$ & $\bar{N} N $ & $\bar{N} E_+ +\bar{E}_+ N $ & $\bar{N} E_- +\bar{E}_- N $ \\
\hline
set-1 & $170$ & $3.68\times 10^{-1}$ & $1.19 \times 10^{-1}$ & $5.20\times 10^{-2}$\\
set-1 & $230$ & $1.21\times 10^{-1}$ & $6.90\times 10^{-2}$ & $2.89\times 10^{-2}$\\
set-2 & $460$ & $7.76\times 10^{-3}$ & $3.04\times 10^{-3}$ & $1.04\times 10^{-3}$\\
set-2 & $540$ & $3.87\times 10^{-3}$ & $2.17\times 10^{-3}$ & $7.24\times 10^{-4}$\\
set-3 & $980$ & $1.93\times 10^{-4}$ & $9.28\times 10^{-5}$ & $2.69\times 10^{-5}$\\
set-3 & $1020$ & $1.53\times 10^{-4}$ & $8.29\times 10^{-5}$ & $2.37\times 10^{-5}$\\
\hline
\end{tabular}
\caption{The production cross sections(in pb) at the LHC14  for  one generation in scenario-B in our model. The masses of $N$ are in the unit of GeV. }
\label{tab:LHC_EN_production}
\end{center}
\end{table}

\section{Conclusions}

An anomaly-free gauged $\Uel$ lepton model was constructed to study the nature of lepton number.
Different from previous studies in the literature, we found two solutions which are free of the anomaly for each SM fermion generation. Our solutions also do not require type-I seesaw mechanism for
active neutrino mass generation.
The price we pay is introducing four extra chiral fermion fields per generation.
While the two solutions whose anomaly cancelation is nontrivial, look superficially similar. The fermion content of one solution is displayed in Tab.\ref{tb:lA}.
We have constructed the minimal scalar sector, Tab.\ref{tb:sA}, such that the active neutrino masses are generated radiatively
without significant fine-tuning the model parameters.  Moreover, the new leptons acquire their masses from the vacuum expectation values of SM Higgs doublet and the scalars, $\phi_{1,2}$, which carry nonzero lepton numbers.

An immediate phenomenological consequence is the existence of a new gauge boson, $Z_\eul$, which is universal for any gauged $U(1)_\ell$ model.
The mass of $Z_\eul$, $M_X$, is determined by the lepton charges of the scalars $\phi_{1,2}$ and the lepton number violating VEVs, $\langle\phi_{1,2}\rangle$.
The $Z_\eul$ boson interferes with the SM photon and $Z$-boson in the process $e^+e^-\ra l\bar{l}$.
Even if the center-of-mass energy of the $e^+e^-$ collider is below the mass of $Z_\ell$, its effects
can be unambiguously identified from the $Z$-line shape and the forward-backward asymmetries. This can be seen in Fig.\ref{fig:Z_LS} and Fig.\ref{fig:AFB}. In contrast, the front-back asymmetries of the quarks will not change from their SM
values. The combination of these two measurements will shed light on the nature of any extra $Z$ boson.

As noted previously, $Z_\eul$ can be produced at a hadron collider by radiating
from a lepton line from the usual Drell-Yan process via the reaction $pp\ra \ell\bar{\ell}\ell^\prime
\bar{\ell}^\prime$ with the invariant mass of one pair of leptons, $\ell \bar{\ell}^\prime$, say, peaking
at $M_X$. The final states to look for are four leptons with no jets.
We found that the LHC14, LHC30, and LHC100 with an integrated luminosity $3000(fb)^{-1}$ can probe the lepton-number violating scale up to roughly
$0.5, 1.0$, and $2.0$ TeV, respectively. This is the same range of direct $Z_\eul$ production can be reached at the $e^+ e^-$ colliders such as ILC500 and CLIC at 2 TeV. The latter also provides much cleaner environment.
For the constraint on the coupling for a light scenario-(B)-type $Z_\ell$,  see \cite{CSKim}.

 Since the SM fermions content is anomalous under $U(1)_\ell$, new heavy leptons are mandated to
cancel the anomaly for a UV-complete theory. The masses of these exotic leptons are usually free parameters and can be heavier than the reach of any foreseeable colliders. We studied the
phenomenologically more interesting case where their masses are  $< 1.0$ TeV. The production of the heavy charged lepton pair at an $e^+ e^-$ collider is $ \sim O(10-10^2) fb $, and the signals for their detection
 can be very clean. Due to the negligible mixing of $E,N$ with their SM counter parts, the usual
 detection channels do not apply. For example, for a heavy neutrino $N$, the usual detection channel
 is $N\ra e W$. However, for our solution, $N$ predominantly decays into final states with $W + 3 \ell$
 (see Eqs.(\ref{eq:NWE}),(\ref{eq:NSe} )). We have used the result that $h_1(a_1)$ will have sizable
 couplings to SM charged leptons with $Y_1\gtrsim 0.1$.

 The production and detection at a hadron collider is much more complicated. While the production cross sections are not too small at the LHC, i.e. $ {\cal O}( 1- 10^2) fb$  for $M_E,M_N< 500$ GeV, one needs a comprehensive study of the SM backgrounds for each possible final state. For a heavy neutrino with the usual decay this has been
 extensively studied before\cite{NPP}. Our preferred final states are different
 and typically involve multi-leptons and  no accompany jet activities other than hadronic $W$. We shall leave such a comprehensive study to future work.
 We note that multi-lepton signals at the LHC were investigated in \cite{AguilarSaavedra, CD, Ssusy} for various scenarios and different models.

Moreover, we have also studied the imprints of the new scalars and heavy leptons at low energies.
It is found that the most prominent constraint on the new fields is from the oblique parameters, $\triangle S$ and $\triangle T$. We have carefully studied the current experimental bounds on  $\triangle S$ and $\triangle T$ and the direct searches for the new charged scalar and heavy charged lepton as well.
The electroweak precision bounds require that the new heavy leptons have to be nearly degenerate.
In any gauged $U(1)_\eul$ model with the custodial symmetry, the approximate mass degeneracy of the new  $SU(2)$ doublet leptons is a generic feature. Depending on their masses, the mass splitting between the heavy neutrino and charged lepton has to be less than ${\cal O}(1- 10\%)$. This will severely constrain the parameters of the model. We look forward to future improvements on the measurements of these quantities.

\begin{acknowledgments}
 WFC is supported by the Taiwan Ministry of Science and Technology under
Grant Nos. 106-2112-M-007-009-MY3  and 105-2112-M-007-029. WFC is thankful for the hospitality of Prof. Chi-Ting Shih and the Department of Applied Physics of Tunghai University where part of his work was done.
TRIUMF receives federal funding via a contribution agreement with the National Research Council of Canada and the Natural Science and Engineering Research
Council of Canada.
\end{acknowledgments}

\bibliographystyle{apsrev4-1}
\bibliography{ref_U1lep}

\end{document}